\documentclass[a4paper,12pt]{article}
\usepackage{amsfonts}
\usepackage{amsthm}
\usepackage{multirow}   
\usepackage{setspace}
\usepackage{amssymb}
\usepackage{amsmath}
\usepackage{graphicx,color,epstopdf}
\usepackage{hyperref}
\usepackage{xcolor}
\usepackage{booktabs}
\usepackage{subfigure}
\usepackage{array}

\usepackage{dsfont }  
\usepackage{algorithm}  
\usepackage{algorithmic}
\usepackage{enumerate}     
\usepackage{multirow}
\usepackage[sort&compress, authoryear]{natbib}
\usepackage{setspace}

\usepackage{float}

\usepackage{graphicx}

\newtheorem{theorem}{Theorem}[section]

\newtheorem{remark}{Remark}[section]
\newtheorem{definition}{Definition}[section]

\allowdisplaybreaks[4]
\newtheorem{assumption}{Assumption}[section]

\numberwithin{equation}{section}

\begin{document}

	\title{
Functional Change Point Detection via Adjacent Deviation Subspace

	}

	\author{Luoyao Yu$^{1}$, Long Feng$^{2}$ and Xuehu Zhu$^{1*}$\\
		$^1$ School of Mathematics and Statistics, Xi'an Jiaotong University, Xi'an, China\\
	$^2$ School of Statistics and Data Science, KLMDASR, LEBPS and LPMC,\\
 Nankai University, Tianjin, China}

	\maketitle
	\begin{abstract}
	
This paper develops the concept of the Adjacent Deviation Subspace (ADS),  a novel framework for reducing  infinite-dimensional functional data into finite-dimensional vector or scalar representations while preserving critical information of functional change points.
To identify this functional subspace, we propose an efficient dimension reduction operator  that overcomes the critical limitation of information loss inherent in traditional functional principal component analysis (FPCA). 
Building upon this foundation,  we first construct a test statistic based on the dimension-reducing target operator to test the existence of change points. Second, we present the MPULSE criterion to estimate change point locations in lower-dimensional representations. This approach not only reduces computational complexity and mitigates false positives but also provides intuitive graphical visualization of change point locations. 
Lastly, we establish a unified analytical framework that seamlessly integrates dimension reduction with precise change point detection.
Extensive simulation studies and real-world data applications validate the robustness and efficacy of our method, consistently demonstrating superior performance over existing approaches.

\end{abstract}


\section{Introduction}

Functional data, characterized by observations recorded as functions rather than scalars or vectors, has established itself as a pivotal domain in modern statistics. The foundations of functional data analysis (FDA) were laid in the 1980s-1990s through seminal works \citep{ramsay1982data, ramsay1991some}, { and the subsequent three decades have seen exponential growth in both theoretical frameworks and methodological applications \citep{yao2005functional,Ramsay2005, muller2008functional,yao2010functional,muller2011functional,hsing2015theoretical, wang2016functional}.
	Extending the classical dimension reduction framework of Principal Component Analysis (PCA) to functional settings, Functional Principal Component Analysis (FPCA) serves as a cornerstone technique in FDA \citep{yao2003shrinkage,yao2006penalized,hall2006properties, yao2007functional,chiou2014multivariate, chen2015localized, happ2018multivariate}.} It projects infinite-dimensional functional observations onto finite-dimensional subspaces, thereby facilitating the application of conventional multivariate statistical methods to functional data. However, while FPCA provides a dimension reduction framework for FDA, its dimension reduction process inherently incurs information loss due to the truncation of components orthogonal to the principal directions.


Change point detection, a statistical methodology designed to identify abrupt structural shifts in data, traces its theoretical foundations to the seminal contributions of \cite{page1954continuous} and \cite{page1955test}. Over seven decades of methodological development have propelled this technique to become an essential analytical tool spanning diverse domains, ranging from corporate finance \citep{bardwell2019most}, U.S. macroeconomic data \citep{stock2012disentangling}, image analysis \citep{tianming2022hyperspectral}, electroencephalogram (EEG) interpretation \citep{kirch2015detection}, air pollution assessment \citep{gagliardi2021change}.
Common change point detection methods include Cumulative Sum (CUSUM) statistics \citep{page1954continuous}, Moving Sum (MOSUM) statistics \citep{bauer1980an}, and non-parametric approaches \citep{matteson2014a}. These methods have been extensively applied to detect change points in scalar or vector-valued data. Their theoretical and methodological developments are well-documented in studies such as \cite{Shao2010}, \cite{killick2012optimal}, \cite{fryzlewicz2014wild}, \cite{kcp2019}, and \cite{zou2020consistent}. For comprehensive reviews of this field, refer to \cite{niu2016multiple} and \cite{aue2024state}.

In recent years, functional change point detection, which aims to identify structural breaks in functional data, has garnered increasing attention \citep{berkes2009detecting, aston2012detecting, aue2018detecting, li2022functional, jiao2023break}. 
Despite substantial progress, two critical challenges persist in this domain. 
First, within the framework of functional change points, FPCA continues to be the most widely used technique. However, the inherent susceptibility of FPCA to information loss may compromise the structural fidelity of change points.
Second, most existing research has predominantly focused on the At Most One Change (AMOC) problem, whereas these methods designed for detecting multiple change points frequently encounter challenges such as high false positive rates and substantial computational complexity.

In this paper, we propose an adaptive dimension reduction framework specifically designed to preserve critical information while addressing the inherent limitations of information loss in conventional FPCA. 
We first formalize the concept of the Adjacent Deviation Subspace (ADS), which theoretically ensures that the number and locations of change points remain identical before and after dimension reduction.
Next, we innovatively develop the efficient dimension reduction operator to identify this functional subspace. 
Moreover, we reveal critical limitations of classical FPCA, particularly highlighting its failure to retain information about structural breaks.  Additionally, we derive the convergence rate and asymptotic distribution of the target operator, providing rigorous theoretical foundations for the proposed framework. 
Based on the asymptotic distribution of our proposed dimension-reduction target operator, we construct an ADS-based statistic for change point testing. 
To maximize power while controlling the Type I error rate, we employ a data-splitting strategy to select the optimal projection directions. This approach divides the data into two parts, where one part is used to determine the best projection directions and the other part is for conducting statistical inference
\citep{Meinshausen2006, Wasserman2009, Cai2024}.
Importantly, our test is not restricted to single change point alternatives but can also handle multiple change points, making it more broadly applicable. To estimate the locations of change point in reduced-dimensional data, motivated from \cite{zhao2020detecting}, which focuses on univariate change detection,  we propose a Multivariate PULSE (MPULSE) criterion.  Our approach not only  flexibly adapts to multivariate settings, but also inherits the advantages of the PULSE criterion. While alleviating computational complexity and addressing false positives,  its visualization capability enables the intuitive identification of change point locations through graphical plots.
By integrating dimension reduction with the ADS-based test and the MPULSE criterion, we establish a comprehensive adaptive framework for functional change point analysis. This framework addresses two critical challenges: (1) preserving information during dimension reduction, and (2) enabling simultaneous testing and estimation of multiple change points.



The remainder of this paper is structured as follows. Section 2 introduces the foundational framework of the adjacent deviation subspace and presents its estimation procedure. Section 3 presents an ADS-based statistic for change point testing. Section 4 proposes a MPULSE criterion for change point estimation and rigorously investigates its consistency properties. Section 5 conducts extensive simulation studies to assess empirical performance, and Section 6 demonstrates practical applications using two real-world datasets. Section 7 discusses the methodological strengths, inherent limitations, and potential directions for future research. All detailed proofs of theoretical results are provided in the Supplementary Material.

\section{Adjacent Deviation Subspace (ADS)}\label{ADS}

\subsection{Notation and Problem Setup}

The following notations are used throughout the paper. Let $\mathcal{T}$ denote a compact subset of $\mathbb{R}$. Let $x(t)$ and $y(t)$ be two square-integrable $\mathbb{R}$-valued functions defined on $\mathcal{T}$. Their inner product is defined as $\langle x(t), y(t) \rangle = \int_{\mathcal{T}} x(t) y(t) \, dt$, where we use the shorthand notation $\int$ in place of $\int_{\mathcal{T}}$. The $L^2$-norm of $x(t)$ is given by $\| x(t) \| = \sqrt{\langle x(t), x(t) \rangle}$. 
Let $ A $ be an operator from $ L^2(\mathcal{T}) $ to $ L^2(\mathcal{T}) $ with the kernel function $ A(t_1,t_2) $. The square of its Hilbert-Schmidt (HS) norm is defined as the integral of the square of the kernel function: 
$\| A \|_{HS}^2 = \int \int |A(t_1,t_2)|^2 \, dt_1 \, dt_2.$ If the HS norm of $ A $ is finite, then it is called a Hilbert-Schmidt operator. 
For any $ x(t) \in L^2(\mathcal{T}) $, we define $A(x)(t)= \int A(t,t_1)x(t_1)dt_1$.
The tensor product $ x(t) \otimes y(t) $ is an operator defined by  $ ( x \otimes y ) (z)(t) = \int x(t) y(t_1)z(t_1)dt_1 $ for all $ z(t) \in L^2(\mathcal{T}) $. Equivalently, for any $ t_1, t_2 \in \mathcal{T} $, the kernel of $ x(t) \otimes y(t) $ satisfies $ (x \otimes y)(t_1, t_2) = x(t_1) y(t_2) $.
Let $ \operatorname{ran}(A) $ denote the range of $ A $, and $\overline{\operatorname{ran}}(A) $ its closure. The notations `$\stackrel{\mathcal{D}}{\to}$' and `$\stackrel{p}{\to}$' denote `convergence in distribution' and `convergence in probability', respectively.

Consider independent $\mathbb{R}$-valued observations $Y_i(t)$ satisfying the model:
\begin{equation}\label{model}
	Y_i(t)=\mu_i(t)+\epsilon_i(t), \quad t \in \mathcal{T}, \quad 1 \leq i \leq n,
\end{equation}
where $\mu_i(t) = E(Y_i(t))$ is the mean function, and $\epsilon_i(t)$ represents a random fluctuation with $E(\epsilon_i(t)) = 0$. 
Without losing generality, we assume that for $1 \leq i \leq n$, $Y_i(t)$, $\mu_i(t)$, and $\epsilon_i(t)$ are square integrable functions. 
We define the covariance operator $\Sigma_i$ with the kernel $\Sigma_i (t_1,t_2)= E\{\epsilon_i (t_1) \epsilon_i (t_2)  \}$ for $i = 1, \dots, n.$


We are interested in detecting changes in the mean function $ \mu_i(t) $. Suppose that the functional sequence $\{\mu_i (t) \}_{i=1}^n$ follows a piecewise structure with $K+1$ segments.  Thus, there are $K$ change points
$1 \leq z_1 < z_2 <...< z_K \leq n$ such that
\begin{equation}\label{1}
	\mu_{z_{k-1}+j}(t) =\mu^{(k)}(t), \quad  \Sigma_{z_{k-1}+j} (t_1,t_2)=\Sigma^{(k)}(t_1,t_2) \quad {\rm and} \quad \mu^{(k)}(t) \neq \mu^{(k+1)}(t),
\end{equation}
for $k=1, \dots, K$, and $1\leq j\leq z_{k}-z_{k-1}$, with  $z_0=0$ and $z_{K+1}=n$. For $1 \leq i \leq K+1$, we assume that $Y_{z_{i-1}+1}(t),\dots,Y_{z_{i}}(t)$ are i.i.d. samples of $Y^{(i)} (t)$  from the stochastic process $P^{(i)}$. It implies that $Y^{(i)}(t) = \mu^{(i)}(t) + \epsilon^{(i)}(t)$ with the covariance operator $\Sigma^{(i)}$. The kernel of $\Sigma^{(i)}$ is given by $\Sigma^{(i)} (t_1,t_2)= E\{\epsilon^{(i)} (t_1) \epsilon^{(i)} (t_2)  \}$.
Let $n_k$ denote the length between two consecutive change points for $k = 1, \dots, K+1$, that is, $n_k = z_k - z_{k-1}$, for $i = 1, \dots, K+1$. Throughout this paper, we assume that $n_i/n \to c_i > 0$.

\subsection{Concept  of ADS}

For the model (\ref{model}), our primary focus is on determining whether the adjacent observations share the same means. The pivotal quantity of interest is the adjacent-segment contrast 
$\mu^{(k+1)}(t)-\mu^{(k)}(t)$, for $k=1,\dots, K$, which motivates the definition of the following functional subspace.

\begin{definition}\label{define1}
	${\rm{Span}}\{\mu^{(k+1)}(t)-\mu^{(k)}(t), \ {\rm{for}} \ k= 1,2,\dots,K \}$ is called the Adjacent Deviation Subspace (ADS) of the sequence $\{Y_i(t)\}_{i=1}^n$ and  is written as $\mathbb{S}_{\{E(Y_i(t))\}_{i=1}^n}$. For this functional subspace, $q={\rm{dim}}\{\mathbb{S}_{\{E(Y_i(t))\}_{i=1}^n}\}$ is called the structural dimension of  $\mathbb{S}_{\{E(Y_i(t))\}_{i=1}^n}$.
\end{definition}

According to Definition \ref{define1}, it is straightforward to observe that $q \le K$. For any basis functions $\{v_1,v_2,...,v_{q}\}$ of  $\mathbb{S}_{\{E(Y_i(t))\}_{i=1}^n}$, we define the $q$-dimensional random vectors $f(Y_i(t))=(f_1(Y_i(t)),\dots,f_q(Y_i(t)))$ with $f_j(Y_i(t))= \langle Y_i(t), v_j(t)   \rangle $ for $j=1,\dots,q$ and $i=1,\dots,n$. Then we have the following theorem.

\begin{theorem}\label{the1}
	Both the functional sequence $\{ Y_i(t) \}_{i=1}^n$ and vector-valued sequence $\{f(Y_i(t))\}_{i=1}^n$ have the same locations of mean changes.
\end{theorem}

\begin{remark}
	Theorem \ref{the1} demonstrates that by applying dimension reduction based on ADS, the infinite-dimensional functional data $\{ Y_i(t) \}_{i=1}^n$ can be reduced to finite-dimensional vector-valued data $\{f(Y_i(t))\}_{i=1}^n$ without losing the information of change points. This advancement overcomes the long-standing limitation of information loss in functional dimension reduction, establishing a rigorous theoretical foundation for applying vector-valued mean change point detection methods to dimension-reduced data.
\end{remark}

\subsection{Estimation of ADS}
Building on the above analysis, our goal turns to identify ADS. Start from the  following ``covariance kernel'' as:
\begin{equation}\label{mn}
	\begin{aligned}
		M_n(t_1,t_2)
		& =\frac{1}{n} \sum_{i=1}^{n} \{ Y_i(t_1)-\bar{Y} (t_1) \}  \{
		Y_i(t_2)-\bar{Y}(t_2)  \},
	\end{aligned}
\end{equation}
where $\bar{Y}(t)=\frac{1}{n}\sum_{i=1}^{n}Y_i(t)$. By calculating the expectation of $M_n(t_1,t_2) $, we find that:
\begin{equation}\label{t1}
	\begin{aligned}
		E\{M_n (t_1,t_2) \}
		&\rightarrow \sum_{j=1}^{K+1} c_j \Sigma^{(j)}(t_1,t_2)+\sum_{i=1}^{K+1} \sum_{j=1}^{K+1} c_ic_j 
		\{ \mu^{(i)}(t_1)-\mu^{(j)}(t_1) \}
		\{\mu^{(i)}(t_2)-\mu^{(j)}(t_2)\}\\	&\equiv:\Sigma_{pooled}(t_1,t_2)+\Delta(t_1,t_2)=M(t_1,t_2).
	\end{aligned}
\end{equation}
Based on (\ref{t1}), we define the operator $ \Delta $ whose kernel is given by:
\begin{equation}\label{delta}
	\Delta(t_1,t_2)=\sum_{i=1}^{K+1} \sum_{j=1}^{K+1} c_ic_j 
	\{ \mu^{(i)}(t_1)-\mu^{(j)}(t_1) \}
	\{\mu^{(i)}(t_2)-\mu^{(j)}(t_2)\}.
\end{equation}

\begin{theorem}\label{deltaZ}
	Under the model (\ref{model}), we have $\overline{\operatorname{ran}}(\Delta) = \mathbb{S}_{\{E(Y_i(t))\}_{i=1}^n}$. Moreover, if $\{v_1(t), \dots, v_q(t)\}$ are the eigenfunctions of $\Delta$ associated with its nonzero eigenvalues, then 
	$\operatorname{Span}\{v_1(t), \dots, v_q(t)\} = \mathbb{S}_{\{E(Y_i(t))\}_{i=1}^n}.$
\end{theorem}

\begin{remark}
	Theorems \ref{the1} and \ref{deltaZ} demonstrate that dimension reduction through the operator $\Delta$ preserves the information of the change points. 
	This provides a methodological foundation for estimating ADS. Meanwhile,  the results in (\ref{t1}) also explains why FPCA sometimes loses information in the change point framework.
\end{remark}

As shown in Theorem \ref{deltaZ}, an accurate estimate of $\Sigma_{pooled}(t_1,t_2)$ is essential to identify the functional subspace $\mathbb{S}_{\{E(Y_i(t))\}_{i=1}^n}$. For the $j$-th segment with $1\leq j \leq K+1$, it can be readily verified that:
$$
E\{ Y_{i+1}(t_1)-Y_{i}(t_1) \} \{Y_{i+1}(t_2)-Y_{i}(t_2) \}= 2 \Sigma^{(j)}(t_1,t_2) \ {\rm{for}}\ z_{j-1}+1 \leq i  \leq z_j-1.  
$$ 
Therefore, we derive the unbiased estimate of $\Sigma^{(j)}(t_1,t_2)$ as:
$$
\Sigma^{(j)}_n(t_1,t_2) =\frac{1}{2(n_j-1)} \sum_{i=z_{j-1}+1}^{z_j-1} \{ Y_{i+1}(t_1)-Y_{i}(t_1) \} \{Y_{i+1}(t_2)-Y_{i}(t_2) \}.
$$
Since the locations of the change points are unknown, we calculate the weighted average of $\Sigma^{(j)}_n(t_1,t_2)$ for $1\leq j \leq K+1$ to get an estimate of $\Sigma_{pooled}(t_1,t_2)$ as:
\begin{equation}\label{sigman}
	\Sigma_{pooled,n}(t_1,t_2)=\frac{1}{2n} \sum_{i=1}^{n-1} \{Y_{i+1}(t_1)-Y_{i}(t_1) \}  \{Y_{i+1}(t_2)-Y_{i}(t_2) \}.
\end{equation}
By combining the equations (\ref{t1}) and (\ref{sigman}), we obtain an estimate of $ \Delta(t_1,t_2) $ as:
\begin{eqnarray}\label{3.5}
	\Delta_n(t_1,t_2)=M_n(t_1,t_2)-\Sigma_{pooled,n}(t_1,t_2).
\end{eqnarray}

%
	
	To investigate the subsequent theoretical properties, we establish the following assumptions.



	\begin{assumption}\label{k2}
		$E\left\| \epsilon^{(i)}(t) \right\|^2< \infty$, for $i=1,\dots, K+1$.
	\end{assumption}

	\begin{assumption}\label{k3}
		$\left\| \mu^{(i)}(t) \right\|^2  < \infty$ , for $i =1,\dots, K+1$.
	\end{assumption}

	\begin{assumption}\label{a3}
		Assume that the nonzero eigenvalues of $ \Delta_n $ and $ \Delta $ are distinct.  
	\end{assumption}

	\begin{assumption}\label{k4}
		$0<\min_{1\leq i\leq n}\lambda_{min}(\Sigma_i)\leq \max_{1\leq i\leq n}\lambda_{max}(\Sigma_i)<\infty.$
	\end{assumption}

	\begin{assumption}\label{k5}
		{\footnotesize
			$0<\min_{1\leq i\leq n-1}\lambda_{min}\big(Var \big( 
			\{2E(Y_{i})(t)-2\bar{E}Y(t)-\epsilon_{i+1}(t)\} \otimes \epsilon_i(t)  \big) \big) \leq \max_{1\leq i\leq n}\lambda_{max}\big( Var \big(  \{2E(Y_{i})(t)-2\bar{E}Y(t)-\epsilon_{i+1}(t) \} \otimes \epsilon_i(t) \big)  \big)<\infty.$
		}
	\end{assumption}

	\begin{assumption}\label{k6}
		$0<\max_{1\leq i\leq n-1}\lambda_{max}\big( \big(E(\{2E(Y_{i})(t)-2\bar{E}Y(t)-\epsilon_{i+1}(t) \} \otimes \epsilon_i(t)  ) \big)^4 \big)<\infty.$
	\end{assumption}	
	
	\begin{assumption}\label{k7}
		$0<\max_{i}\lambda_{max} \big\{E\big( \epsilon_{i+1}(t) \otimes \epsilon_i(t) \big) ^4 \big\} <\infty.$
	\end{assumption}

	\begin{remark}	
		Assumption \ref{k2} ensures the boundedness of the second-order moment of the residual term, which is required for the consistency of variance estimation. Assumption \ref{k3} ensures $\Sigma^{(1)},\dots,\Sigma^{(s+1)}$ and $\Sigma_{pooled}$ are well-defined,  and further guarantee that $\Delta$ is a Hilbert-Schmidt operator. 
		Assumption \ref{a3} ensures the uniqueness of eigenvalues and eigenfunctions, thereby guaranteeing the existence of their asymptotic distribution, as referenced in \cite{Anderson1963} and \cite{hsing2015theoretical}. 
		Assumptions~\ref{k4}, \ref{k5}, \ref{k6}, and \ref{k7} are about the existence of moments  and are satisfied in many situations, such as  the cases where $\Sigma_i$ is an identity operator and $m$-dependent.
		These assumptions are imposed to satisfy the Lindeberg condition, thereby enabling the application of the central limit theorem for deriving the asymptotic distribution of $\Delta_n$.
		These assumptions are also commonly used, see relevant references  \cite{hsing2015theoretical}, \cite{libing2017} and \cite{li2018}.			
	\end{remark}

	%
	%
	%

The following theorem describes the convergence rate of $ \Delta_n $.

\begin{theorem}\label{kernel}
	Under Assumptions \ref{k2} and \ref{k3}, we have that $\Delta$ is a Hilbert-Schmidt operator, and 
	$
	\vert\vert\Delta_n-\Delta \vert\vert_{HS}=O_p\left(n^{-1/2} \right).
	$
\end{theorem}

Let $ \hat{v}_j(t) $ and $ v_j(t) $ denote the eigenfunctions corresponding to the eigenvalues $ \hat{\lambda}_j $ and $ \lambda_j $ of $ \Delta_n $ and $ \Delta $, respectively. Let $\bar{E}Y(t)=\sum_{l=1}^{K+1}c_l \mu^{(l)}(t)$ and  
$\Sigma_{\alpha}=\sum_{i=1}^{K+1} c_i Var\bigg(
\langle \alpha(t), \{ (2 \mu^{(i)} -2\bar{E}Y-\epsilon^{(i)} ) \otimes  \epsilon^{(i)} \} (\alpha)(t)  \rangle \bigg).$
Then we have the following limiting distribution.

\begin{theorem}\label{dis}
	Suppose that Assumptions \ref{k2}-\ref{k6} hold.  For any function $\alpha(t) \in L^2(\mathcal{T}) $  satisfying $\vert\vert\alpha(t)\vert\vert=1$, we have:
	$$
	\sqrt{n} \langle \alpha(t), (\Delta_n-\Delta)(\alpha)(t) \rangle \stackrel{\mathcal{D}}{\to}
	\mathcal{N}\left(0,\Sigma_{\alpha}   \right).
	$$	
	Furthermore, we have
	$$
	\begin{aligned}
		&\sqrt{n} (\hat{\lambda}_j-\lambda_j)
		\stackrel{\mathcal{D}}{\to}
		\langle v_j(t), \tilde{\Delta}(v_j)(t) \rangle \  and   \
		\sqrt{n} \{ \hat{v}_j(t)-v_j(t)  \} \stackrel{\mathcal{D}}{\to}  
		\sum_{\lambda_k \neq \lambda_j}  \frac{1}{\lambda_j-\lambda_k}  P_k 	\tilde{\Delta} (v_j)(t),
	\end{aligned}
	$$
	for $j=1,\dots, q$,	where $ \langle \alpha(t), \tilde{\Delta} ( \alpha) (t)\rangle$ follows 
	the norm distribution
	$\mathcal{N}\left(0,\Sigma_{\alpha}    \right).$		
\end{theorem}

\begin{remark} Theorem \ref{kernel} demonstrates that $\Delta_n$ is $\sqrt{n}$-consistent for
	$\Delta$. Furthermore, Theorem \ref{dis} establishes the limiting distribution of $\Delta_n$ and its related eigenvalue problem. 
	These theoretical results will serve in subsequent change point testing and estimation, as detailed in Sections \ref{test} and \ref{estimation}.
\end{remark}

Let $\{\phi_i(t), i\geq 1  \}$ be an orthonormal basis of $ L^2(\mathcal{T}) $. 
Suppose the following approximation holds:
\begin{equation}\label{b1}
	Y_i(t) = C_i^{\top} \phi(t), \quad \text{and} \quad 	\hat{v}_j(t) = B_j^{\top} \phi(t),
\end{equation}
where  $C_i = (c_{i1}, \dots, c_{iD})^{\top}$, $B_j = (b_{j1}, \dots, b_{jD})^{\top}$, and $\phi(t) = \left( \phi_1(t), \phi_2(t), \dots, \phi_D(t) \right)^{\top}$. We assume that $D$ is chosen such that the approximation closely approximates the original functions. Such approximations are common in FDA \citep{Ramsay2005,wang2016functional}. Since $\phi(t)$ are basis functions, it is easy to see that $\int \phi(t) \phi^{\top}(t)  dt$ is an identity matrix.
Therefore, $\Delta_n (\hat{v}_j)(t)  $ can be transformed as follows:
\begin{equation*}
	\begin{aligned}
		\Delta_n (\hat{v}_j)(t) 
		=  &\int \Delta_n(t,t_1) \hat{v}_j(t_1)dt_1 \\
		=&\int \frac{1}{n} \sum_{i=1}^{n} \{Y_i(t)-\bar{Y}(t) \} \{ Y_i(t_1)-\bar{Y}(t_1) \} \hat{v}_j(t_1)dt_1 \\
		&- \int \frac{1}{2n} \sum_{i=1}^{n-1} \{Y_{i+1}(t)-Y_{i}(t) \} \{Y_{i+1}(t_1)-Y_{i}(t_1) \} \hat{v}_j(t_1)dt_1 \\
		=& \phi^{\top}(t) \bigg\{    \frac{1}{n} \sum_{i=1}^{n}  (C_i-\bar{C}) (C_i-\bar{C})^{\top} - \frac{1}{2n} \sum_{i=1}^{n-1}   (C_{i+1}-C_{i}) (C_{i+1}-C_{i})^{\top} \bigg\} B_j\\         
		=& \phi^{\top}(t) A_n  B_j, 
	\end{aligned}
\end{equation*} 
where $\bar{Y}(t) = \bar{C}^{\top} \phi(t)$ with $\bar{C}= \frac{1}{n} \sum_{i=1}^{n} C_i$, and 
\begin{equation}\label{cn}
	A_n= \frac{1}{n} \sum_{i=1}^{n}  (C_i-\bar{C}) (C_i-\bar{C})^{\top} - \frac{1}{2n} \sum_{i=1}^{n-1}   (C_{i+1}-C_{i}) (C_{i+1}-C_{i})^{\top} .
\end{equation} 
Based on the above discussion, we have the following inferences:
\begin{equation*}
	\begin{aligned}
		\Delta_n (\hat{v}_j)(t)= {\lambda}_j \hat{v}_j(t)  
		& \Rightarrow \phi^{\top}(t) A_n B_j = {\lambda}_j  {\phi}^{\top}(t)B_j.  
	\end{aligned}
\end{equation*} 
Because $\phi(t)$ are basis functions,  its coefficients must necessarily be equal, that is 
\begin{equation*}
	A_n  B_j ={\lambda}_j B_j,
\end{equation*} 
which implies that $B_j$ is an eigenvector corresponding to $A_n$. Then the eigenfunction $\hat{v}_j(t)$ can be derived in (\ref{b1}). Thus, the reduced data $\hat{f}_j(Y_i(t))$ can be obtained as follows:
\begin{equation}\label{fn}
	\hat{f}_j(Y_i(t))= \langle Y_i(t), \hat{v}_j(t)   \rangle 
	= C_i^{\top}   B_j,
\end{equation}
for $i = 1, 2, \dots, n$ and $j = 1, \dots, q$.

\subsection{ Dimension Estimation of ADS}
As commented above, determining the structural dimension $q$ is a crucial step to identify the functional subspace $\mathbb{S}_{\{E(Y_i(t))\}_{i=1}^n}$. 
Inspired from \cite{zhu2020Dimensionality}, we employ a thresholding ridge ratio (TRR) criterion to estimate  $ q $ as:
\begin{eqnarray}\label{TRR}
	\hat{q}:=\max_{1\leq k \leq p-1}\left\{k: \  \frac{\hat{\lambda}_{k+1}+c_n}{\hat{\lambda}_k+c_n} \leq \tau_1 \right\},
\end{eqnarray}
where the ridge value $c_n$  tends to zero at a proper rate and the thresholding value $\tau_1$ satisfies $0<\tau_1 <1$.
In accordance with the recommendation provided in \cite{zhu2020Dimensionality}, selecting 
$\tau_1=0.5$  is justified as it effectively mitigates the risk of overestimation typically associated with larger values of 
$\tau_1$, while simultaneously avoiding the underestimation that often arises with smaller values of  $\tau_1$.
Because the target matrix involved herein is different from those in \cite{zhu2020Dimensionality} and there is no theoretical result for optimal selection of the ridge value $c_n$, we recommend choosing $c_n=0.5\log (\log(n)) /\sqrt{n}$ according to the numerical studies. The algorithm is very easy to implement and the consistency of $\hat{q}$ is stated in the following theorem.
\begin{theorem}\label{con-hatq}
	Under the same conditions in Theorem~\ref{kernel}, if $c_n$ satisfies $c_n\to 0$ and $\sqrt{n} c_n \to \infty$ as $n\to \infty$, then $P\left(\hat{q}=q\right)\to 1$.
\end{theorem}

\section{ADS-based Test for Change Points }\label{test}

In this section,  we check whether the functional sequence $\{\mu_i(t)\}_{i=1}^n$ contains structural shifts. 
\subsection{The Test Statistic Construction}

We reformulate the null hypothesis as:
\begin{eqnarray}\label{H0}
	H_0: \mu_1(t)=\mu_2(t)=\cdots=\mu_n(t),
\end{eqnarray}
against the alternative hypothesis that the sequence of parameters $\{\mu_i(t)\}_{i=1}^n$ changes, that
\begin{eqnarray}\label{H1}
	H_1: \exists 1 \leq z_1 < z_2 < \cdots < z_K < n, \quad \mu_i(t) =
	\begin{cases}
		\mu^{(1)}(t), & 0   <i \leq z_1, \\
		\mu^{(2)}(t), & z_1 < i \leq z_2, \\
		\vdots & \\
		\mu^{(K+1)}(t), & z_K < i \leq n,
	\end{cases}
\end{eqnarray}
where $ \mu^{(1)}(t), \mu^{(2)}(t), \dots, \mu^{(K+1)}(t) $ represent the mean functions in each segment, with  $ \mu^{(k)}(t) \neq \mu^{(k+1)}(t) $ for $k=1,\dots,K$, and $K \geq 1$.

Based on the ADS framework in Section 2, under the null,  the operator $ \Delta $ whose kernel is defined in (\ref{delta}) equals zero, and its estimate $\Delta_n$ given in (\ref{3.5}) satisfies that  for any fixed function $\alpha \in L^2(\mathcal{T}) $  satisfying $\vert\vert\alpha\vert\vert=1$,  $$
\sqrt{n}\langle \alpha(t), \Delta_n(\alpha)(t) \rangle\stackrel{\mathcal{D}}{\to}
\mathcal{N}\left(0,\sigma_{\alpha}^4   \right),
$$
where  $\sigma_{\alpha}^2=Var (\langle   \epsilon_i(t), \alpha(t) \rangle)$. Adopting the similar estimator in (\ref{sigman}), $\sigma_{\alpha}^2$ can be consistently estimated by 
\begin{equation}\label{sig}
	\hat{\sigma}_{\alpha}^2=\frac{1}{2n}  \sum_{i=1}^{n-1} \langle \alpha(t), Y_{i+1}(t)-Y_{i}(t)  \rangle^2.
\end{equation}
Therefore, for any fixed function $\alpha \in L^2(\mathcal{T}) $  satisfying $\vert\vert\alpha\vert\vert=1$, we construct a test statistic as:
\begin{equation}\label{Stand_test}
	T_n(\alpha(t))=\sqrt{n} V_n(\alpha(t)) / \hat{\sigma}_{\alpha}^2.
\end{equation}
Then we have following asymptotic properties.
\begin{theorem}\label{null_alter}
	Suppose  Assumptions \ref{k2}, \ref{k3}, \ref{a3}, \ref{k4} and \ref{k7} hold.  For any function $\alpha(t) \in L^2(\mathcal{T}) $  satisfying $\vert\vert\alpha(t)\vert\vert=1$, we have:
	\begin{itemize}
		\item Under $H_0$,
		$$
		T_n (\alpha(t)) \stackrel{\mathcal{D}}{\to}
		\mathcal{N}\left(0,1 \right);
		$$
		\item Under $H_1$, 	$$
		T_n(\alpha(t))/n^{1/2} \stackrel{p}{\to} \langle \alpha(t), \Delta (\alpha)(t)  \rangle \left\{ \sum_{i=1}^{K+1} c_i  (\sigma^{(i)}_{\alpha} )^2 \right\} ^{-1},
		$$
		where $(\sigma^{(i)}_{\alpha})^2  =Var (\langle  \epsilon^{(i)}(t), \alpha(t) \rangle)$ for $i=1,\dots,K+1$.
	\end{itemize}
\end{theorem}

Based on Theorem  \ref{null_alter}, we derive the final statistic $T_n$ via maximizing the power as:
\begin{equation}\label{Stand_test}
	T_n=\sqrt{n} V_n(\alpha^{\star}(t)) / \hat{\sigma}_{\alpha^{\star}}^2,
\end{equation}
where 
\begin{equation}\label{as}
	\alpha^{\star}(t)=  \mathop{\arg\max}_{\alpha \in L^2(\mathcal{T}), \ \vert\vert\alpha\vert\vert=1} \langle \alpha(t), \Delta (\alpha)(t)  \rangle \left\{ \sum_{i=1}^{K+1} c_i  (\sigma^{(i)}_{\alpha} )^2 \right\} ^{-1}.
\end{equation}	

To evaluate the sensitivity of the proposed test statistic to the alternative hypothesis, we assume a local alternative framework with a single change point. Specifically, we consider the following sequence of local alternative models:
\begin{eqnarray}\label{local}
	H_{1n} : \mu^{(2)}(t) = \mu^{(1)}(t) + C_n \delta_0(t),
\end{eqnarray}
where $C_n \rightarrow 0$ as $n \rightarrow \infty$, and $\delta_0(t)$ is a nonzero function without depending on the sample size $n$. It is worth noting that when  $C_n$ is fixed,  $H_{1n}$ reduces to the global alternative hypothesis $H_1$. The following theorem presents the power performance of the test statistic under the global and local alternative hypotheses.

\begin{theorem}\label{local1}
	Suppose Assumptions \ref{k2}, \ref{k3}, \ref{a3}, \ref{k4}, and \ref{k7} hold. Then, we have:
	\begin{itemize}
		\item Under the global alternative $H_1$, 
		$$
		T_n/n^{1/2} \stackrel{p}{\to} \langle \alpha^{\star}(t), \Delta (\alpha^{\star})(t)  \rangle \left \{ \sum_{i=1}^{K+1} c_i  (\sigma^{(i)}_{\alpha^{\star}} )^2 \right\} ^{-1},
		$$
		\item Under the local alternative $H_{1n}$ with $C_n=n^{-1/4} $, 
		$$
		T_n\stackrel{\mathcal{D}}{\to} \mathcal{N}(\gamma_1,1),
		$$
		where 
		$$
		\gamma_1= \langle \alpha^{\star}(t), \delta_0(t) \rangle^2 \left \{  c_1  (\sigma^{(1)}_{\alpha^{\star}} )^2 +c_2  (\sigma^{(2)}_{\alpha^{\star}} )^2   \right \} ^{-1}.
		$$

	\end{itemize}

\end{theorem}

\begin{remark}
	Theorem \ref{null_alter} demonstrates that under the null hypothesis, the test satistic  $T_n$ converges in distribution to $\mathcal{N}(0,1)$. 
	This ensures that the critical value can be easily determined without relying on Monte Carlo approximation or resampling techniques.
	On the other hand, Theorem \ref{local1} indicates that  $T_n$ can detect the local alternatives distinct from the null at the rate of order $n^{-1/4}$. To construct a powerful test, it is necessary and important to determine an optimal choice of projection direction. Next, we will present a data-splitting strategy for estimating $\alpha^{\star}(t)$ in Subsection \ref{date_splitting}.   
\end{remark}

\subsection{Optimal Projection Direction Selection: Data-splitting Strategy}\label{date_splitting}

Based on the definition of $T_n$, the optimal projection direction  $\alpha^{\star}(t)$ satisfies the equation (\ref{as}) and maximizes power.
To ensure Type I error control while optimizing power, we employ a data-splitting strategy, which guarantees independence between the selected projection direction and the data used to construct the statistic.

Generally, data splitting involves randomly dividing the data into two groups. In change point analysis, however, the data inherently has an ordered structure, making random splitting unsuitable as it may lose change point information. Therefore, inspired by \cite{zou2020consistent}, we adopt the following specific data splitting strategy for change point problems.
For convenience, we assume the sample size is even, i.e. $n=2T$.
Given the dataset $\mathcal{D} = \{Y_1(t), \dots, Y_n(t)\}$, we divide the sample into two subsets:
$$
\mathcal{D}_1 = \{Y_1(t), Y_3(t), \dots, Y_{2T-1}(t)\} \quad \text{and} \quad \mathcal{D}_2 = \{Y_2(t), Y_4(t), \dots, Y_{2T}(t)\}.
$$
Based on (\ref{3.5}),  we can obtain the estimates $\Delta_{1n}$ and $\Delta_{2n}$ through $\mathcal{D}_1$ and through $\mathcal{D}_2$, respectively.
Then we adopt $\mathcal{D}_1$ to estimate the projection direction $\alpha_{1n}(t)$ as:
\begin{equation}\label{a1}
	\alpha_{1n}(t)= \mathop{\arg\max}_{\alpha \in L^2(\mathcal{T}), \ \vert\vert\alpha\vert\vert=1} \frac{\sqrt{T}  \langle \alpha(t), \Delta_{1n}(\alpha)(t) \rangle}{ \hat{\sigma}_{\alpha}^2},
\end{equation}
where $ \hat{\sigma}_{\alpha}^2=\frac{1}{2T}  \sum_{Y_i \in  \mathcal{D}_1 } \langle \alpha(t), Y_{i+1}(t)-Y_{i}(t)  \rangle^2 .$

Next, we  conduct the test statistic $T_{2n}$ depended on $\mathcal{D}_2$ as:
\begin{equation}\label{tn2}
	T_{2n}=
	\frac{\sqrt{T}  \langle \alpha_{1n}(t), \Delta_{2n}(\alpha_{1n})(t) \rangle}{ \hat{\sigma}_{\alpha_{1n}}^2 }, 
\end{equation}
where 
$\hat{\sigma}_{\alpha_{1n}}^2=
\frac{1}{2T}  \sum_{Y_i \in  \mathcal{D}_2 } \langle \alpha_{1n}(t), Y_{i+1}(t)-Y_{i}(t)  \rangle^2.$

The following theorem describes the theoretical properties of $T_{2n}$.

\begin{theorem}\label{t2n}
	If Assumptions \ref{k2}, \ref{k3}, \ref{a3}, \ref{k4}, and \ref{k7} hold,  we have the following results.
	\begin{itemize}
		\item  Under $H_0$,
		$$
		T_{2n} \stackrel{\mathcal{D}}{\to} \mathcal{N}(0,1).
		$$
		\item Under the global alternative hypotheses $H_1$, $$T_{2n}/T^{1/2}  \stackrel{p}{\to}  \langle \alpha^{\star}(t), \Delta (\alpha^{\star})(t)  \rangle \left\{ \sum_{i=1}^{K+1} c_i  (\sigma^{(i)}_{\alpha^{\star}} )^2 \right\} ^{-1}.$$
		\item Under the local alternative hypotheses $H_{1n}$,
		$$T_{2n} \stackrel{\mathcal{D}}{\to} \mathcal{N}(\gamma_1,1),$$ 
		where  $\gamma_1= \langle \alpha^{\star}(t), \delta_0(t) \rangle^2  \{  c_1  (\sigma^{(1)}_{\alpha^{\star}} )^2 +c_2  (\sigma^{(2)}_{\alpha^{\star}  } )^2    \} ^{-1}.$
	\end{itemize}

\end{theorem}

\begin{remark}
	Theorem \ref{t2n} establishes the validity of the test statistic $T_{2n}$ for hypothesis testing within the data splitting framework. This method effectively controls the Type I error rate while maintaining high power. Therefore, it is reasonable to adopt the data splitting approach to select the projection direction $ \alpha^{\star}(t) $. This approach is well-documented in the literature \citep{Meinshausen2006, Wasserman2009, Cai2024}.
\end{remark}

To calculate $\alpha_{1n}$ in (\ref{a1}) and $T_{2n}$ in (\ref{tn2}), we adopt a similar approximation in (\ref{b1}). Let $\alpha(t) = \alpha_0^{\top} \phi(t)$ and $\alpha_{1n}(t)=\alpha_{0n}^{\top} \phi(t)$ in (\ref{a1}), then we can directly obtain:
\begin{equation}\label{3.10}
	\alpha_{0n}= \mathop{\arg\max}_{\alpha_0 \in \mathbb{R}^D, \ \vert\vert\alpha_0\vert\vert=1} \frac{\sqrt{T}  \alpha_0^{\top} A_{1n} \alpha_0 }   { \alpha_0^{\top} Q_{1n} \alpha_0  },
\end{equation}
where 
\begin{equation}\label{an1}
	A_{1n}=\frac{1}{n} \sum_{Y_i \in \mathcal{D}_1 }  (C_i-\bar{C}) (C_i-\bar{C})^{\top} - \frac{1}{2n} \sum_{Y_i \in \mathcal{D}_1}   (C_{i+1}-C_{i}) (C_{i+1}-C_{i})^{\top},
\end{equation}
and 
\begin{equation}\label{bn1}
	Q_{1n}=\frac{1}{2n} \sum_{Y_i \in \mathcal{D}_1}   (C_{i+1}-C_{i}) (C_{i+1}-C_{i})^{\top}.
\end{equation}
Setting $\alpha_0 = Q_{1n}^{-1/2} \beta_0$ and $\alpha_{0n}=Q_{1n}^{-1/2} \beta_{0n}$, 
the optimization problem in (\ref{3.10}) reduces to solving the following problem:
$$
\beta_{0n}= \mathop{\arg\max}_{\beta_0 \in \mathbb{R}^D, \ \vert\vert\beta_0\vert\vert=1} \sqrt{T} \beta_0^\top Q_{1n}^{-1/2} A_{1n} Q_{1n}^{-1/2} \beta_0,
$$
which implies that $\beta_{0n}$ is the eigenvector corresponding to the largest eigenvalue of $Q_{1n}^{-1/2} A_{1n} Q_{1n}^{-1/2}$, which consequently gives $\alpha_{0n} = Q_{1n}^{-1/2} \beta_{0n}$ and $ \alpha_{1n}(t) = \alpha_{0n}^\top \phi(t) $. 

Next, using $ \mathcal{D}_2 $, we calculate the test statistic as:  
\begin{equation}\label{t2}
	T_{2n} = \frac{\sqrt{T} \, \alpha_{0n}^\top A_{2n} \alpha_{0n} }{\alpha_{0n}^\top Q_{2n} \alpha_{0n}},
\end{equation}
where 
\begin{equation}\label{an2}
	A_{2n}=\frac{1}{n} \sum_{Y_i \in \mathcal{D}_2 }  (C_i-\bar{C}) (C_i-\bar{C})^{\top} - \frac{1}{2n} \sum_{Y_i \in \mathcal{D}_2}   (C_{i+1}-C_{i}) (C_{i+1}-C_{i})^{\top},
\end{equation}
and 
\begin{equation}\label{bn2}
	Q_{2n}=\frac{1}{2n} \sum_{Y_i \in \mathcal{D}_2}   (C_{i+1}-C_{i}) (C_{i+1}-C_{i})^{\top}.
\end{equation}
Based on the above description, we formulate the complete Algorithm \ref{algorithm1}.

\begin{algorithm}[htb!]
	\caption{ADS-based test for functional change point testing}
	\label{algorithm1}
	\begin{algorithmic}[1]
		\REQUIRE $\mathcal{D} = \{Y_1(t), \dots, Y_n(t)\}$;
		\STATE Partition $\mathcal{D}$ into two disjoint subsets $\mathcal{D}_1=\{Y_1(t), Y_3(t), \dots, Y_{2T-1}(t)\}$, $\mathcal{D}_2=\{Y_2(t), Y_4(t), \dots, Y_{2T}(t)\}$;
		\STATE Calculate $A_{1n}$ and $Q_{1n}$ based on (\ref{an1}), (\ref{bn1}), and $A_{2n}$ and $Q_{2n}$ based on (\ref{an2}), (\ref{bn2});
		\STATE Obtain the estimate $\beta_{0}$ of the eigenvector corresponding to the largest eigenvalue of $Q_{1n}^{-1/2} A_{1n} Q_{1n}^{-1/2}$, then get $\alpha_{0n} = Q_{1n}^{-1/2} \beta_{0n}$;
		\STATE Calculate $T_{2n}$ based on (\ref{t2}).
		\ENSURE Reject $H_0$, if $|T_{2n}|> z_{\alpha/2}$ where $z_{\alpha/2}$ is the upper $\alpha/2$ quantile of standard normal distribution $N(0,1)$.\\
	\end{algorithmic}
\end{algorithm}

\section{Estimate Locations of Change Points: MPULSE Criterion}\label{estimation}

In this section, motivated by \citep{zhao2020detecting}, we propose a Multivariate PULSE (MPULSE) criterion to estimate the locations of multiple change points in dimension-reduced sequence. Specifically, we define the following statistic $S_{n}(i)$ based on the ridge ratios:
\begin{equation}\label{D1}
	S_{n}(i)=\min_{1\leq l\leq \hat{q}}\left\{{{|\tilde D_{l,n}(i)|+\tilde{c}_{n}} \over {|\tilde D_{l,n}(i+{3\over2}\tilde{\alpha}_n)|}+\tilde{c}_{n}}\right\},
\end{equation}
where
\begin{equation}\label{D2}
	\tilde D_{l,n}(i)={1\over \tilde{\alpha}_n}\sum_{j=i}^{i+\tilde{\alpha}_n-1}D_{l,n}(j) \ {\rm{with}}\ D_{l,n}(i)={1\over \tilde{\alpha}_n}\left(  \sum_{j=i}^{i+\tilde{\alpha}_n-1}
	\hat{f}_j(Y_i(t))
	-\sum_{j=i+\tilde{\alpha}_n}^{i+2\tilde{\alpha}_n-1}
	\hat{f}_j(Y_i(t)) \right),
\end{equation}
and $\tilde{c}_{n}$ is a ridge parameter with a small value to avoid unstable $0/0$ terms.  Here we suggest using $\tilde{\alpha}_n=\lfloor n^{0.6} \rfloor$ and $\tilde{c}_n=0.25\sqrt{\log n\over \tilde{\alpha}_n}$ as proposed in \cite{zhao2020detecting}.

\begin{figure}[htb!]
	\centering
	\includegraphics[width=7cm,height=4cm]{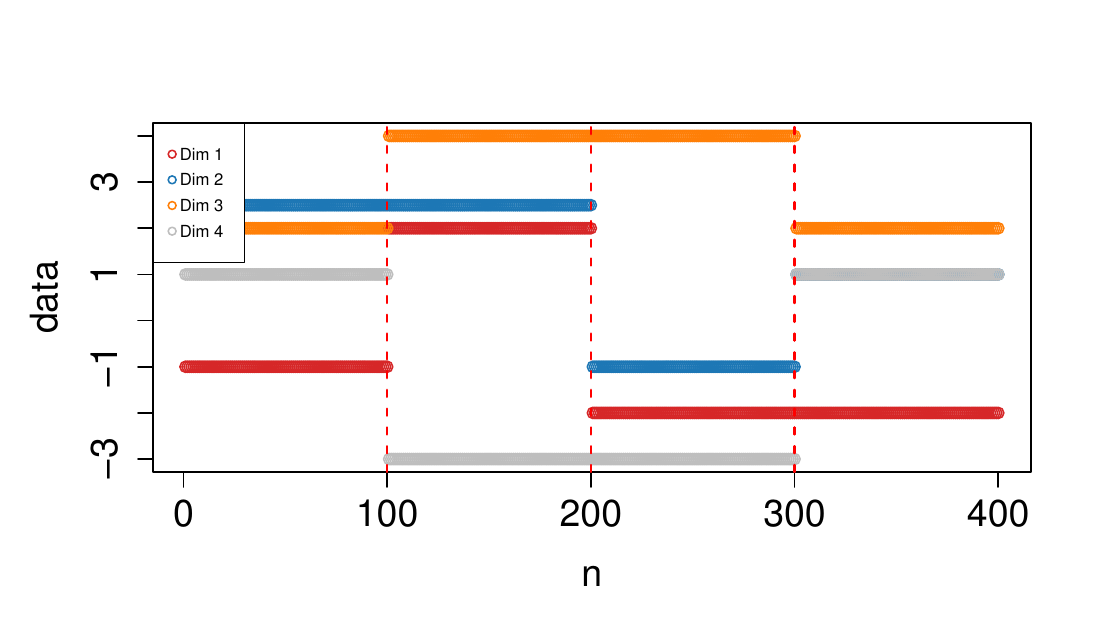}
	\centering
	\includegraphics[width=7cm,height=4cm]{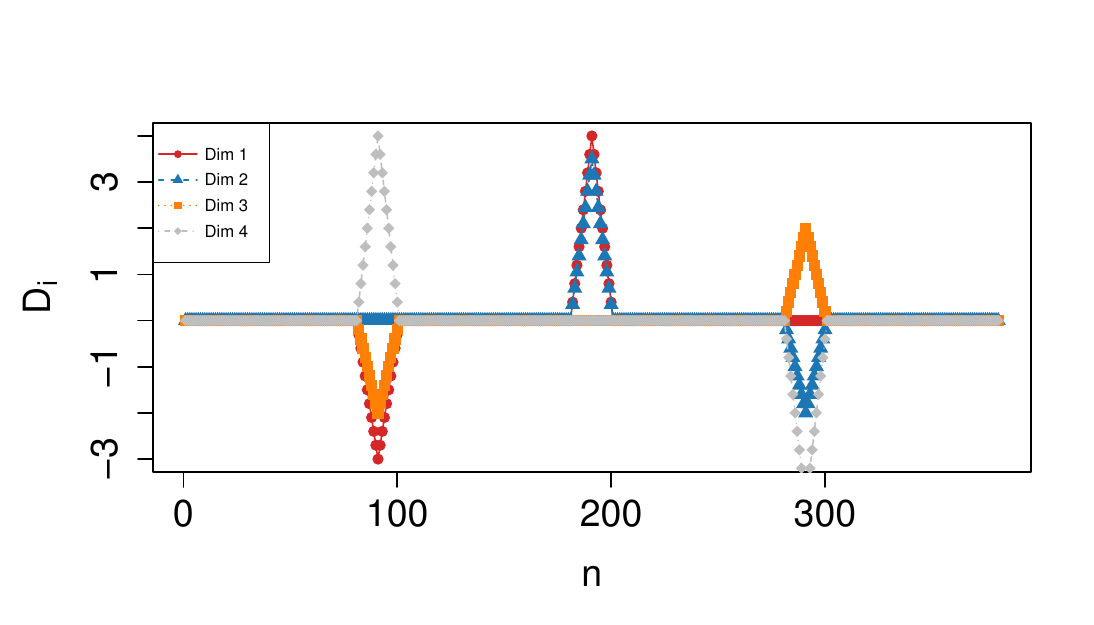}
	
	\centering
	\includegraphics[width=7cm,height=4cm]{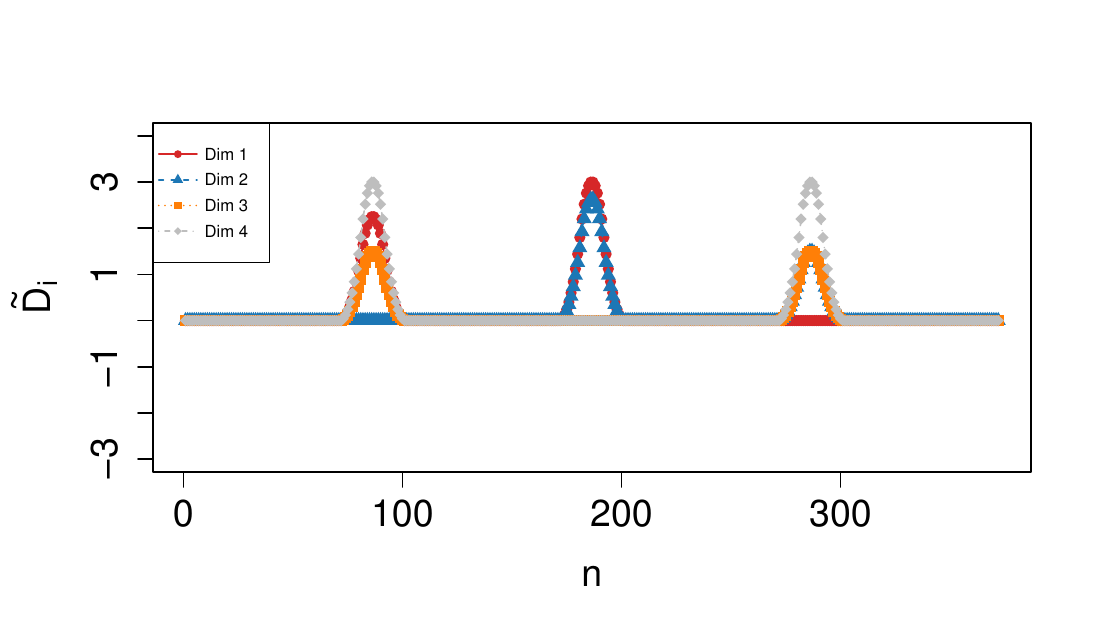}
	\centering
	\includegraphics[width=7cm,height=4cm]{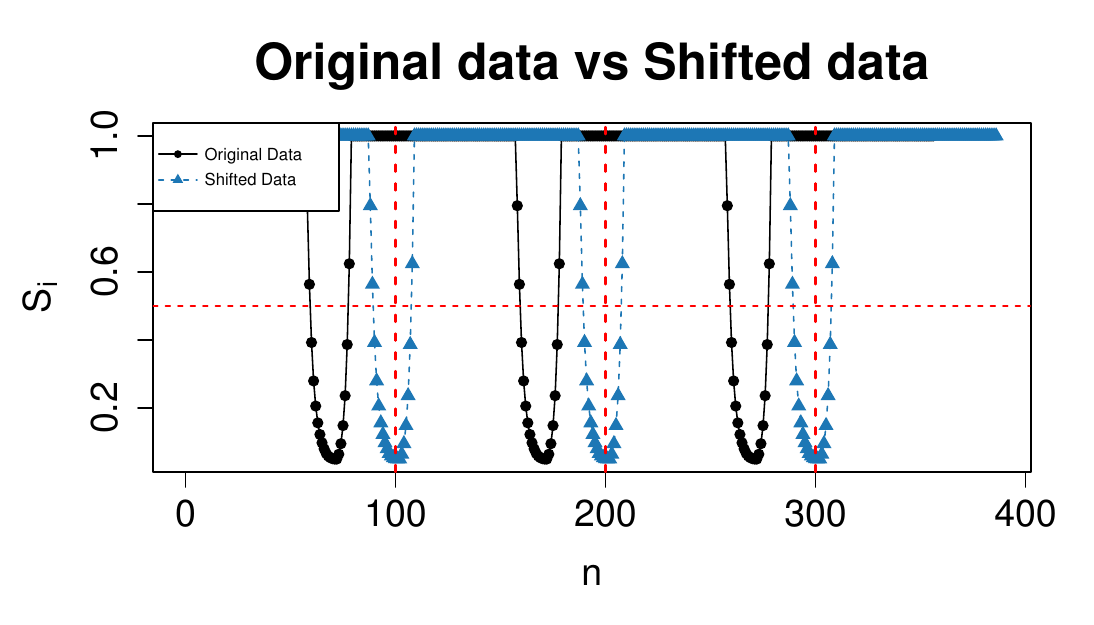}
	\caption{The plots at the population level.}\label{fig1}
\end{figure}

Figure \ref{fig1} visualizes the conceptual framework of MPULSE at the population level. It shows that while both $\tilde D_{l,n}$ and $D_{l,n}$ exhibit jumps at change points, they face challenges in determining appropriate thresholds. In contrast, the local minima of $S_{n}$ consistently equal zero, significantly simplifying the threshold selection. Furthermore, we observe that change points in different dimensions are effectively consolidated into $S_{n}$. By identifying the local minima of $S_{n}$, we can accurately estimate multiple change points in  multivariate scenarios. Based on the above analysis, we establish the following criterion using a threshold parameter.

We constrain the search region to guarantee that each selected interval contains exactly one local minimum of $S_{n}(i)$. This is implemented through a threshold parameter $\tau_2$ ($0<\tau_2<1$) with the following criterion:
\begin{eqnarray}\label{TMAR}
	\{ i_l, 1 \leq i \leq n-{7\over 2}\tilde{\alpha}_n: S_{n}(i)< \tau_2\}.
\end{eqnarray}
To simplify the search process, we maintain the recommended value $\tau_2=0.5$ as proposed in \cite{zhao2020detecting}.

Additionally, Figure \ref{fig1} also illustrates that distinct change points correspond to non-overlapping intervals. By applying the thresholding parameter mentioned above, we can identify these disjoint intervals and locate the local minima within each one. Notably, there exists a shift of $3\tilde{\alpha}_n$ between the estimated local minima and the actual change points. Furthermore, at the population level, the separation between any two local minimizers must exceed $2\tilde{\alpha}_n$.
Assume that we identify $\hat{K}$ pairs $(m_{k}, M_{k})$ where $m_{k}$ and $M_{k}$ are determined by the threshold condition $S_{n}(i)< \tau_2$. Specifically, $m_{k}$ satisfies $S_{n}(m_{k}-1)\ge \tau_2$ and $S_{n}(m_{k})< \tau_2$, while $M_{k}$ fulfills $S_{n}(M_{k})< \tau_2$ and $S_{n}(M_{k}+1)\ge  \tau_2$. Within each such interval, we identify 
\begin{equation}\label{i}
	i_k=\mathop{\arg\min}_{m_{k} < i <  M_{k} }S_{n}(i),
\end{equation}	
and it is straightforward to derive that the estimated location of the change point is $\hat z_{k}=i_k+3\tilde{\alpha}_n$. For this estimate, we present the following theoretical guarantees.

\begin{theorem}\label{locations}
	Under the same conditions in Theorem \ref{kernel},  suppose the tuning parameter $\tilde{c}_n$ and the segment length $\tilde{\alpha}_n$ satisfy the following conditions:
	$$
	\frac{\tilde{c}_n \sqrt{\tilde{\alpha}_n}}{\sqrt{\log n}} \to \infty, \quad \frac{n^{1/4} \log n}{\sqrt{\tilde{\alpha}_n}} \to 0, \ {\rm{and}} \   \sqrt{ \frac{\tilde{\alpha}_n \log n}{n}     } \to 0.
	$$
	Then $\hat K= K$ with a probability going to one and the estimators $ \{\hat z_1,..., \hat z_{\hat K}\} $ 
	satisfy that for every $\epsilon>0$, as $n\rightarrow\infty$,
	$$
	\lim_{n \rightarrow \infty} \Pr\left\{ \max_{1\le k\le K}{\Big|{\hat z_k-z_k\over \tilde{\alpha}_n}\Big|}<\epsilon\right\}\rightarrow 1.
	$$
	

\end{theorem}

\begin{remark}
	As a criterion, MPULSE is neither an optimization algorithm nor a hypothesis testing procedure. Therefore, this criterion effectively addresses two common challenges in multiple change point detection: the false positive problem and high computational complexity. 
	
	
\end{remark}

\begin{remark}

	

	Theorem \ref{locations} presents asymptotic results for change point estimation, incorporating the theoretical insights on dimension reduction from Theorem \ref{kernel}. This integration forms a comprehensive framework for functional change point analysis within the context of dimension reduction subspaces, significantly broadening the scope of functional change point analysis.


\end{remark}


Building upon the results obtained in Sections \ref{ADS} and \ref{estimation}, we develop a comprehensive framework for functional change point estimation. The proposed methodology is formally presented in Algorithm \ref{algorithm0}.

\begin{algorithm}[htb]
	\caption{Adaptive functional change point estimation algorithm}
	\label{algorithm0}
	\begin{algorithmic}[1]
		\REQUIRE $Y_1,\dots,Y_n$, take $\tilde{\alpha}_n=\lfloor n^{0.6} \rfloor$, $\tau_1=\tau_2=0.5$,   $c_n=0.5\log (\log(n)) \sqrt{1/n}$,  and $\tilde{c}_n=0.25\sqrt{\log n\over \tilde{\alpha}_n}$;
		\STATE Estimate the target matrix $A_{n}$ in (\ref{cn});
		\STATE Decompose the matrix $A_{n}$ to get  the eigenvalues  $\hat{\lambda}_{1} \geq \dots \geq \hat{\lambda}_{D}$  and the eigenvectors $B_1,\dots, B_D$;
		\STATE Determine the dimension $\hat{q}$ based on TRR in (\ref{TRR});
		\STATE Calculate the low dimensional sequence  $\{\hat{f}(Y_i(t))\}_{i=1}^n$ in (\ref{fn});
		\STATE Compute $D_{l,n}$, $\tilde D_{l,n}$ in (\ref{D2}), and then construct the statistic $\tilde S_{n}(i)$ in (\ref{D1});
		\STATE Get $\hat{K}$ intervals $\{(m_{k}, M_{k})\}_{k=1}^{\hat{K}}$ based on the conditions (\ref{TMAR});
		\STATE Derive $i_k$ in each interval based on (\ref{i}); 
		\STATE Obtain the change point location $\hat z_{k}=i_k+3\tilde{\alpha}_n$;
		\ENSURE $\{\hat z_1,\dots,\hat{z}_{\hat{K}}\}$ and $\hat{K}$.\\
	\end{algorithmic}
\end{algorithm}

\section{Simulations}

In this section, we conduct numerical experiments  to demonstrate the effectiveness of our approaches. The functional data objects are constructed via $D$ Fourier basis functions $\{\phi_d(t)\}_{d=1}^D$ defined on the interval $[0,1]$. Specifically, the basis functions are composed of the following:
\begin{itemize}
	\item A constant function: $\phi_1(t) = 1$;
	\item $(D-1)/2$ cosine functions: $\phi_{2k}(t) = \sqrt{2}\cos(2\pi kt)$ for $k=1,\dots,(D-1)/2$;
	\item $(D-1)/2$ sine functions: $\phi_{2k+1}(t) = \sqrt{2}\sin(2\pi kt)$ for $k=1,\dots,(D-1)/2$,
\end{itemize}
with $D=21$ yielding 10 cosine functions, 10 sine functions, and a constant function.

The sequence $\{C_i\}_{i=1}^n$ is partitioned into $K+1$ segments corresponding to $K$ change points. 
The data is generated by:
\begin{equation*}
	Y_i(t) = C_i^{\top} \phi(t), \ {\rm{for}} \ i=1,\cdots,n,
\end{equation*}
where
\begin{equation*}
	C_i =
	\begin{cases} 
		\mu_D+\epsilon_i, & {\rm{if}}\ z_j+1 \leq  i \leq z_{j+1} \quad \text{and} \quad  j \text{ is even},  \\
		\epsilon_i,  & {\rm{if}}\  z_j+1 \leq  i \leq z_{j+1} \quad \text{and} \quad  j \text{ is odd},
	\end{cases}
\end{equation*}
with the $D$-dimensional mean shift vector $\mu_D = (\underbrace{u,\dots,u}_{D_c},0,\dots,0)^\top$, 
the noise term $\epsilon_i$ is given by $\epsilon_i = \Sigma^{1/2} Z_i$ with $Z_i = (Z_{i1}, \dots, Z_{iD})^{\top}$ and $Z_{il}$ ($l = 1,\dots,D$) being i.i.d. from $G$, the covariance matrix  $\Sigma = (\sigma_{ij})$ is diagonal, with diagonal elements defined by
$\sigma_{ii} = 2^{-i},$ and all off-diagonal elements set to zero.   
To assess the robustness of our method across different distributions, we consider $G=\mathcal{N}(0,1)$ and $t_4$, which represent light-tailed and heavy-tailed distributions, respectively.
It is worth mentioning that this decreasing variance setting is quite common in functional data analyses, used to mimic various decays for the eigenvalues of the covariance operator.
We vary $D_c$, $u$, and $G$ to evaluate the efficacy of all methods at different levels of signal sparsity, shift magnitude, and error structures, respectively.
Meanwhile, to demonstrate the impact of varying numbers of change points on our method, we design the following scenarios:
\begin{itemize}
	\item {\bf Single change point}: The sample size is $n = 200$, and  the change point is located as 100. 
	\item {\bf Multiple change points}: The sample size is $n = 300$, and the two change points are located as 100 and 200, respectively. 	
\end{itemize}

\subsection{Change Point Testing}

In this subsection, we evaluate the finite-sample performance of the proposed ADS-based test. 
We compare our method with two classical functional testing approaches, $T_{ARS}$ \citep{aue2018detecting} and $T_{BGHK}$  \citep{berkes2009detecting}, both implemented in the R package ``fChange''.
Each experiment is repeated 1000 times to compute the empirical size and power.

{\bf Single change point test.} The experimental results, summarized in Table \ref{test1}, demonstrate that $T_{2n}$ effectively controls size while maintaining high power across all scenarios. As the deviation from the null hypothesis decreases, the empirical powers of $T_{ARS}$ and $T_{BGHK}$ drop sharply, sometimes falling below 0.5. Additionally, the heavy-tailed distribution of the error term has no significant impact on the performance of our  test, but substantially degrades the performance of  $T_{ARS}$ and $T_{BGHK}$. Specifically, under the configuration $G=t_4$, $D_c=10$, $u=0.06$, $T_{ARS}$ achieves a power of merely 0.143, $T_{BGHK}$ reaches only 0.231, whereas $T_{2n}$ reaches 0.818. Overall, $T_{2n}$ performs significantly better than $T_{ARS}$ and $T_{BGHK}$ for single change point testing.

{\bf Multiple change points test.} The empirical sizes and powers of the multiple change point test are summarized in Table \ref{test2}.  Our findings indicate that $T_{2n}$ maintains strong performance in both empirical size and power. In contrast, $T_{ARS}$ and $T_{BGHK}$ performs poorly in detecting multiple change point alternatives.  Specially, when the deviation from the null hypothesis is small, which corresponds to $u=0.06$, $T_{ARS}$ and $T_{BGHK}$ fails to detect the alternative hypotheses and their empirical powers are below 0.2. Conversely, our proposed test successfully detects these alternatives, achieving empirical power exceeding $0.92$. Across all scenarios, $T_{2n}$ demonstrates robust performance regardless of the distribution $G$, shift magnitude $u$, signal sparsity $D_c$, or the number of change points $K$.

\begin{table}[htp!]
	\caption{Empirical sizes and powers of the tests for single change point}
	\centering
	\begin{tabular}{ccc|ccc|ccc}
		\hline
		&            & \multicolumn{1}{l}{} & \multicolumn{3}{c}{$D_c=10$}  & \multicolumn{3}{c}{$D_c=20$}  \\ \hline
		$G$                       & method     & $u=0$                & $u=0.06$ & $u=0.08$ & $u=0.1$ & $u=0.06$ & $u=0.08$ & $u=0.1$ \\  \hline
		\multirow{3}{*}{$N(0,1)$} & $T_{2n}$      & 0.055                & 0.996    & 1.000    & 1.000   & 1.000    & 1.000    & 1.000   \\
		& $T_{ARS}$  & 0.072                & 0.250    & 0.579    & 0.929   & 0.718    & 1.000    & 1.000   \\
		& $T_{BGHK}$ & 0.076                & 0.451    & 0.708    & 0.919   & 0.539    & 0.875    & 0.994   \\
		\multirow{3}{*}{$t_4$}    & $T_{2n}$      & 0.057                & 0.818    & 0.991    & 1.000   & 1.000    & 1.000    & 1.000   \\
		& $T_{ARS}$  & 0.047                & 0.143    & 0.202    & 0.432   & 0.220    & 0.601    & 0.942   \\
		& $T_{BGHK}$ & 0.066                & 0.231    & 0.381    & 0.622   & 0.267    & 0.513    & 0.782  \\ \hline
	\end{tabular}
	\label{test1}
\end{table}

\begin{table}[htp!]
	\caption{Empirical sizes and powers of the tests for multiple change points}
	\centering
	\begin{tabular}{ccc|ccc|ccc}
		\hline
		&            &       & \multicolumn{3}{c}{$D_c=10$}  & \multicolumn{3}{c}{$D_c=20$}  \\ \hline
		$G$                       & method     & $u=0$ & $u=0.06$ & $u=0.08$ & $u=0.1$ & $u=0.06$ & $u=0.08$ & $u=0.1$ \\  \hline
		\multirow{3}{*}{$N(0,1)$} & $T_{2n}$      & 0.052 & 1.000    & 1.000    & 1.000   & 1.000    & 1.000    & 1.000   \\
		& $T_{ARS}$  & 0.063 & 0.093    & 0.118    & 0.223   & 0.121    & 0.253    & 0.607   \\
		& $T_{BGHK}$ & 0.064 & 0.163    & 0.275    & 0.527   & 0.193    & 0.478    & 0.901   \\
		\multirow{3}{*}{$t_4$}    & $T_{2n}$      & 0.059 & 0.926    & 0.999    & 1.000   & 1.000    & 1.000    & 1.000   \\
		& $T_{ARS}$  & 0.046 & 0.053    & 0.063    & 0.093   & 0.068    & 0.105    & 0.167   \\
		& $T_{BGHK}$ & 0.062 & 0.085    & 0.136    & 0.208   & 0.114    & 0.166    & 0.322  \\ \hline
	\end{tabular}
	\label{test2}
\end{table}

\subsection{Change Point Estimation}

{ 
	
	This subsection evaluates the performance of the MPULSE algorithm. We compare MPULSE with three  classical methods of change point estimation, including E-Divisive \citep{matteson2014a}, Kernel Change Point Algorithm (KCP) \citep{kcp2019}, and Sparsified Binary Segmentation (SBS) \citep{cho2015multiple-change-point}, by implementing them under two dimension reduction frameworks, ADS and FPCA.
	For clarity, we denote the MPULSE versions using these frameworks as MPULSE-ADS (ADS) and MPULSE-FPCA (FPCA), respectively. While SBS is designed for multivariate data, it defaults to Wild Binary Segmentation (WBS) \citep{fryzlewicz2014wild} when the estimated dimension $\hat{q} = 1$. Performance metrics include the average of estimated number of change points $\hat{K}$, root-mean-square error (RMSE) of $\hat{K}$, and Rand index (RI) \citep{rand1971objective} between estimated and true segments. 
	The E-Divisive method, SBS and WBS  are implemented in the R packages ``ecp'', ``hdbinseg'' and ``wbs'', respectively.
	In addition, we compare our method with two functional change point detection methods, the ARS method proposed by \cite{aue2018detecting} and the BGHK method proposed by \cite{berkes2009detecting}, both implemented in the R package ``fChange''. To ensure robustness, all experiments are repeated 100 times.

	{\bf Single change point estimation.} The results of single change point estimaion are presented in Table~\ref{detect2}. The findings are summarized as follows. Under $G \sim \mathcal{N}(0,1)$, both MPULSE-ADS, ARS, and BGHK exhibit  excellent performance, and ADS-based methods consistently outperform FPCA-based approaches in most scenarios.
	Under the heavy-tailed $t_4$ distribution, comparative methods exhibit significant performance degradation due to distributional sensitivity, while MPULSE-ADS maintains accurate and robust change-point detection. 
	Across various scenarios, the results demonstrate the consistent effectiveness of MPULSE-ADS under different conditions, including shift magnitude $ u $, signal sparsity $ D_c $, and distribution $ G $, further highlighting its robustness. 
	
}

\begin{table}[htp!]
	\caption{Results for single change point estimation}
	\centering
	\resizebox{\textwidth}{!}{
		\begin{tabular}{cl|ccc|ccc|ccc|ccc}
			\hline
			\multicolumn{1}{l}{}       &                 & \multicolumn{3}{c}{$D_c=10$, $u=0.08$} & \multicolumn{3}{c}{$D_c=10$, $u=0.1$} & \multicolumn{3}{c}{$D_c=20$, $u=0.08$} & \multicolumn{3}{c}{$D_c=20$, $u=0.1$} \\  \hline
			$G$                        & Method          & $\hat{K}$      & RMSE      & RI        & $\hat{K}$      & RMSE      & RI       & $\hat{K}$      & RMSE      & RI        & $\hat{K}$      & RMSE      & RI       \\  \hline
			\multirow{10}{*}{$N(0,1)$} & MPLUSE-ADS      & 0.95           & 0.61      & 0.78      & 0.97           & 0.54      & 0.81     & 0.96           & 0.42      & 0.86      & 0.97           & 0.30      & 0.91     \\
			& MPLUSE-FPCA     & 1.45           & 0.79      & 0.74      & 1.28           & 0.63      & 0.73     & 1.34           & 0.66      & 0.75      & 1.36           & 0.62      & 0.73     \\
			& KCP-ADS         & 0.44           & 1.23      & 0.59      & 0.93           & 1.64      & 0.64     & 1.68           & 2.14      & 0.73      & 2.41           & 2.58      & 0.80     \\
			& KCP-FPCA        & 0.00           & 1.00      & 0.50      & 0.00           & 1.00      & 0.50     & 0.00           & 1.00      & 0.50      & 0.00           & 1.00      & 0.50     \\
			& E-Divisive-ADS  & 0.63           & 0.81      & 0.70      & 0.74           & 0.66      & 0.77     & 0.93           & 0.59      & 0.85      & 1.00           & 0.37      & 0.92     \\
			& E-Divisive-FPCA & 0.24           & 0.91      & 0.57      & 0.52           & 0.77      & 0.68     & 0.76           & 0.60      & 0.81      & 1.06           & 0.24      & 0.98     \\
			& SBS-ADS         & 0.78           & 0.86      & 0.70      & 0.82           & 0.72      & 0.77     & 0.93           & 0.61      & 0.84      & 1.08           & 0.57      & 0.93     \\
			& SBS-FPCA        & 0.42           & 0.76      & 0.68      & 0.77           & 0.48      & 0.85     & 0.98           & 0.14      & 0.96      & 1.00           & 0.00      & 0.99     \\
			& ARS             & 0.64           & 0.60      & 0.77      & 0.92           & 0.28      & 0.92     & 1.00           & 0.00      & 0.98      & 1.00           & 0.00      & 0.99     \\
			& BGHK            & 0.72           & 0.53      & 0.81      & 0.90           & 0.32      & 0.91     & 0.85           & 0.39      & 0.87      & 1.00           & 0.00      & 0.98     \\  \hline \hline
			\multirow{10}{*}{$t_4$}    & MPLUSE-ADS      & 1.15           & 0.67      & 0.78      & 1.13           & 0.66      & 0.79     & 1.15           & 0.69      & 0.80      & 1.21           & 0.57      & 0.86     \\
			& MPLUSE-FPCA     & 1.43           & 0.81      & 0.75      & 1.38           & 0.68      & 0.75     & 1.52           & 0.76      & 0.74      & 1.45           & 0.70      & 0.74     \\
			& KCP-ADS         & 0.05           & 0.98      & 0.51      & 0.09           & 0.95      & 0.54     & 0.66           & 1.50      & 0.59      & 0.75           & 1.35      & 0.68     \\
			& KCP-FPCA        & 0.00           & 1.00      & 0.50      & 0.00           & 1.00      & 0.50     & 0.00           & 1.00      & 0.50      & 0.00           & 1.00      & 0.50     \\
			& E-Divisive-ADS  & 0.50           & 0.85      & 0.62      & 0.57           & 0.79      & 0.67     & 0.70           & 0.68      & 0.75      & 0.95           & 0.61      & 0.85     \\
			& E-Divisive-FPCA & 0.18           & 0.95      & 0.54      & 0.23           & 0.89      & 0.58     & 0.21           & 0.92      & 0.56      & 0.75           & 0.69      & 0.78     \\
			& SBS-ADS         & 2.10           & 2.24      & 0.65      & 1.86           & 1.90      & 0.71     & 1.99           & 1.82      & 0.76      & 2.38           & 2.34      & 0.82     \\
			& SBS-FPCA        & 0.03           & 0.98      & 0.51      & 0.09           & 0.95      & 0.54     & 0.25           & 0.87      & 0.61      & 0.73           & 0.52      & 0.85     \\
			& ARS             & 0.22           & 0.88      & 0.59      & 0.49           & 0.71      & 0.71     & 0.59           & 0.64      & 0.76      & 0.92           & 0.28      & 0.93     \\
			& BGHK            & 0.41           & 0.77      & 0.67      & 0.54           & 0.68      & 0.72     & 0.45           & 0.74      & 0.68      & 0.80           & 0.45      & 0.84   \\ \hline 
		\end{tabular}
	}
	\label{detect2}
\end{table}

{ 
	
	{\bf Multiple change point estimation.} The results shown in Table \ref{detect4} demonstrate consistent performance patterns under different distributions. Under Gaussian conditions, MPULSE-ADS maintains superior performance compared to alternative methods, whereas ARS and BGHK prove ineffective for multiple change point estimation. ADS-based approaches consistently outperform FPCA-based methods. Most notably, under heavy-tailed $t_4$ distributions, conventional methods suffer substantial performance deterioration, whereas MPULSE-ADS preserves both estimation accuracy and robustness.
	Comprehensive simulation results confirm that our proposed MPULSE-ADS delivers superior detection performance and exceptional robustness, demonstrating consistent effectiveness across all evaluated conditions and influencing factors.

}

\begin{table}[htp!]
	\caption{Results for multiple change point estimation}
	\centering
	\resizebox{\textwidth}{!}{
		\begin{tabular}{cl|ccc|ccc|ccc|ccc}
			\hline
			\multicolumn{1}{l}{}       &                 & \multicolumn{3}{c}{$D_c=10$, $u=0.08$} & \multicolumn{3}{c}{$D_c=10$, $u=0.1$} & \multicolumn{3}{c}{$D_c=20$, $u=0.08$} & \multicolumn{3}{c}{$D_c=20$, $u=0.1$} \\ \hline
			$G$                        & Method          & $\hat{K}$      & RMSE      & RI        & $\hat{K}$      & RMSE      & RI       & $\hat{K}$      & RMSE      & RI        & $\hat{K}$      & RMSE      & RI       \\  \hline
			\multirow{10}{*}{$N(0,1)$} & MPLUSE-ADS      & 1.43           & 0.88      & 0.78      & 1.58           & 0.66      & 0.81     & 1.81           & 0.48      & 0.86      & 1.98           & 0.20      & 0.92     \\
			& MPLUSE-FPCA     & 1.78           & 0.80      & 0.73      & 2.00           & 0.65      & 0.77     & 2.10           & 0.57      & 0.79      & 2.03           & 0.44      & 0.80     \\
			& KCP-ADS         & 1.00           & 2.77      & 0.42      & 2.76           & 3.72      & 0.60     & 4.34           & 4.74      & 0.69      & 6.37           & 5.71      & 0.81     \\
			& KCP-FPCA        & 0.00           & 2.00      & 0.33      & 0.00           & 2.00      & 0.33     & 0.00           & 2.00      & 0.33      & 0.00           & 2.00      & 0.33     \\
			& E-Divisive-ADS  & 0.84           & 1.51      & 0.57      & 1.52           & 1.06      & 0.77     & 1.74           & 0.85      & 0.85      & 2.08           & 0.37      & 0.96     \\
			& E-Divisive-FPCA & 0.14           & 1.91      & 0.38      & 0.46           & 1.73      & 0.47     & 0.84           & 1.52      & 0.58      & 2.04           & 0.20      & 0.98     \\
			& SBS-ADS         & 0.70           & 1.64      & 0.52      & 1.46           & 1.12      & 0.75     & 1.70           & 0.95      & 0.84      & 2.01           & 0.41      & 0.96     \\
			& SBS-FPCA        & 0.03           & 1.98      & 0.34      & 0.20           & 1.89      & 0.40     & 1.06           & 1.37      & 0.67      & 1.86           & 0.53      & 0.94     \\
			& ARS             & 0.12           & 1.91      & 0.38      & 0.20           & 1.84      & 0.41     & 0.19           & 1.85      & 0.41      & 0.53           & 1.55      & 0.56     \\
			& BGHK            & 0.28           & 1.78      & 0.44      & 0.48           & 1.60      & 0.52     & 0.40           & 1.67      & 0.50      & 0.92           & 1.11      & 0.72     \\  \hline \hline
			\multirow{10}{*}{$t_4$}    & MPLUSE-ADS      & 1.64           & 0.86      & 0.76      & 1.76           & 0.62      & 0.80     & 1.80           & 0.62      & 0.81      & 1.81           & 0.48      & 0.87     \\
			& MPLUSE-FPCA     & 2.02           & 0.68      & 0.75      & 1.93           & 0.57      & 0.78     & 2.18           & 0.66      & 0.80      & 1.92           & 0.60      & 0.77     \\
			& KCP-ADS         & 0.24           & 2.08      & 0.37      & 0.39           & 2.26      & 0.38     & 0.92           & 2.62      & 0.43      & 1.80           & 3.04      & 0.57     \\
			& KCP-FPCA        & 0.00           & 2.00      & 0.33      & 0.00           & 2.00      & 0.33     & 0.00           & 2.00      & 0.33      & 0.00           & 2.00      & 0.33     \\
			& E-Divisive-ADS  & 0.60           & 1.66      & 0.50      & 1.00           & 1.46      & 0.60     & 0.97           & 1.40      & 0.62      & 1.60           & 1.03      & 0.80     \\
			& E-Divisive-FPCA & 0.17           & 1.91      & 0.38      & 0.21           & 1.86      & 0.40     & 0.24           & 1.85      & 0.40      & 0.73           & 1.58      & 0.55     \\
			& SBS-ADS         & 2.51           & 2.22      & 0.62      & 2.37           & 2.31      & 0.62     & 2.69           & 2.06      & 0.68      & 3.33           & 2.30      & 0.82     \\
			& SBS-FPCA        & 0.00           & 2.00      & 0.33      & 0.00           & 2.00      & 0.33     & 0.03           & 1.98      & 0.34      & 0.30           & 1.84      & 0.43     \\
			& ARS             & 0.13           & 1.90      & 0.39      & 0.17           & 1.87      & 0.40     & 0.11           & 1.92      & 0.38      & 0.12           & 1.91      & 0.38     \\
			& BGHK            & 0.12           & 1.91      & 0.38      & 0.26           & 1.79      & 0.44     & 0.24           & 1.81      & 0.43      & 0.36           & 1.71      & 0.48   \\ \hline
		\end{tabular}
	}
	\label{detect4}
\end{table}

\begin{figure}[htp!]
	\centering
	\includegraphics[width=7cm,height=3.5cm]{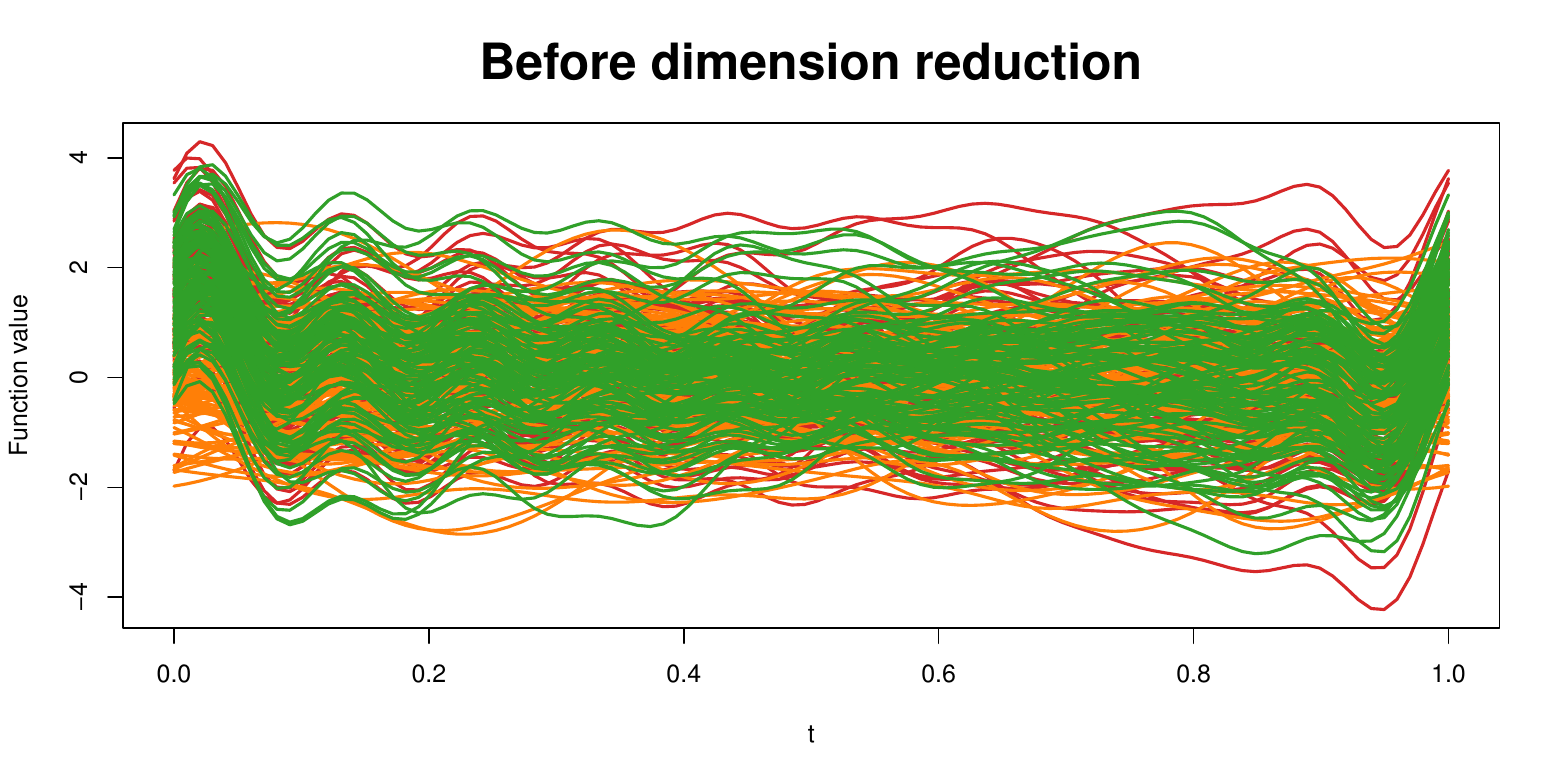}
	\includegraphics[width=7cm,height=3.5cm]{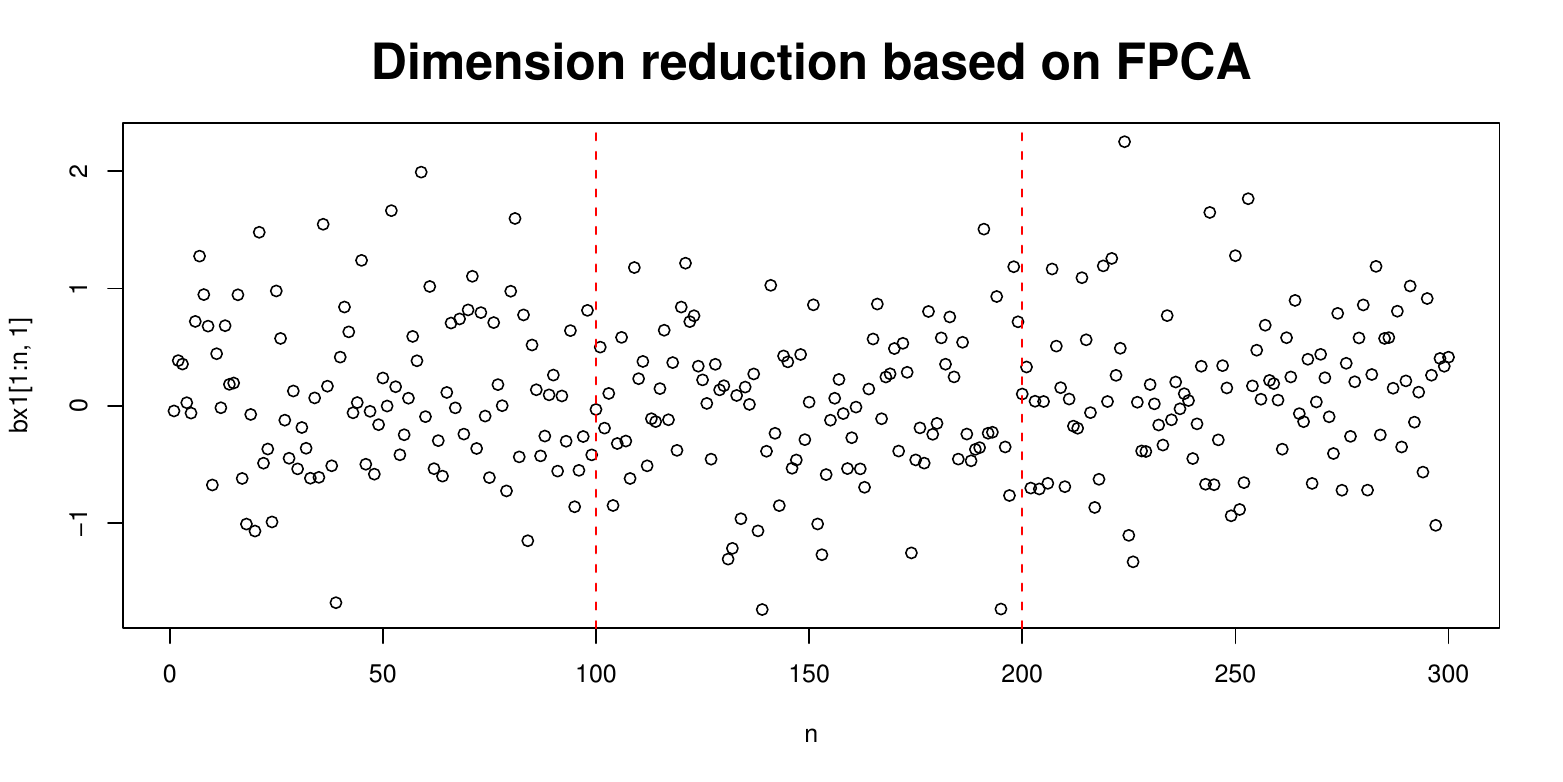}
	\includegraphics[width=7cm,height=3.5cm]{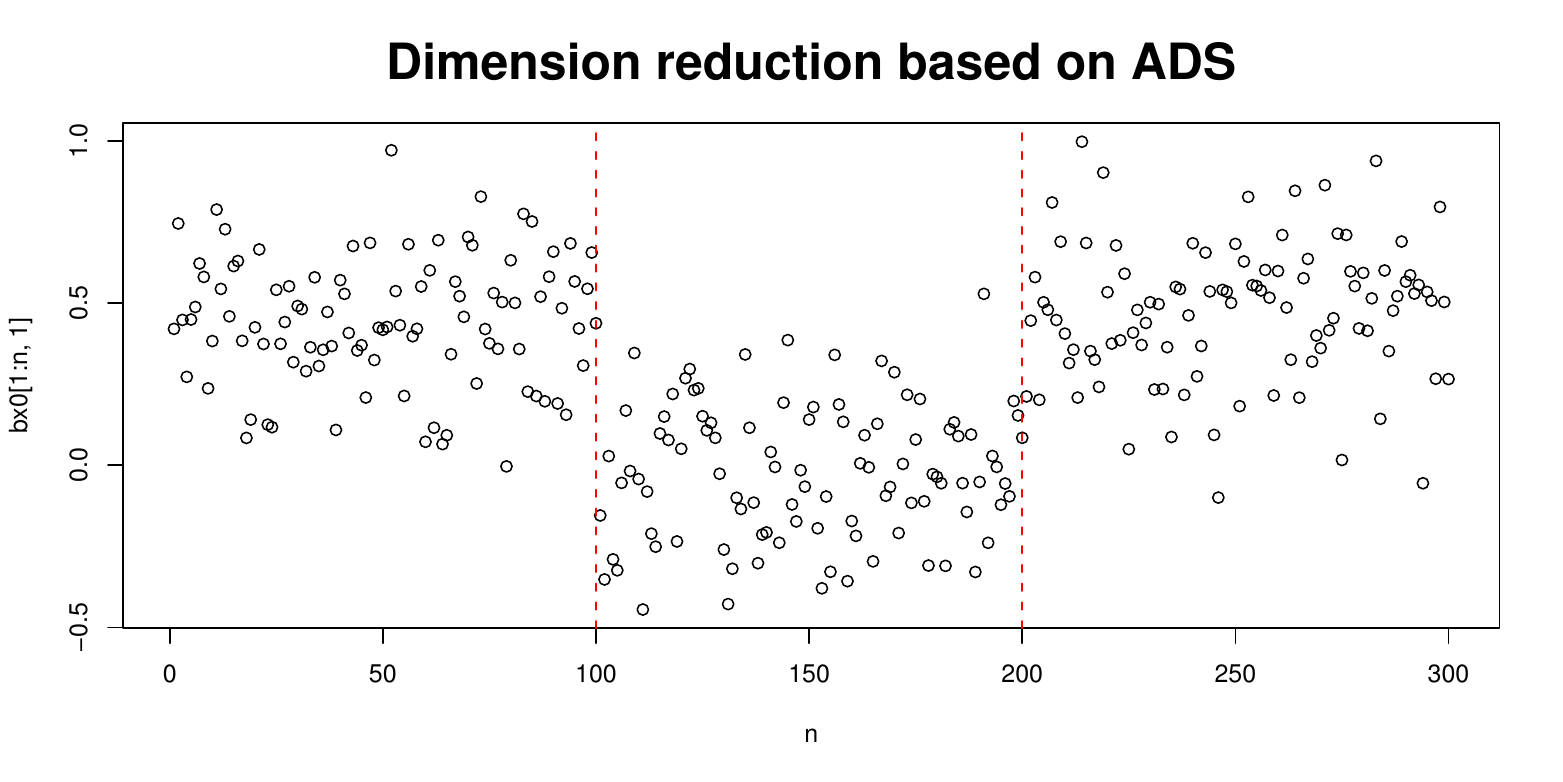}
	\includegraphics[width=7cm,height=3.5cm]{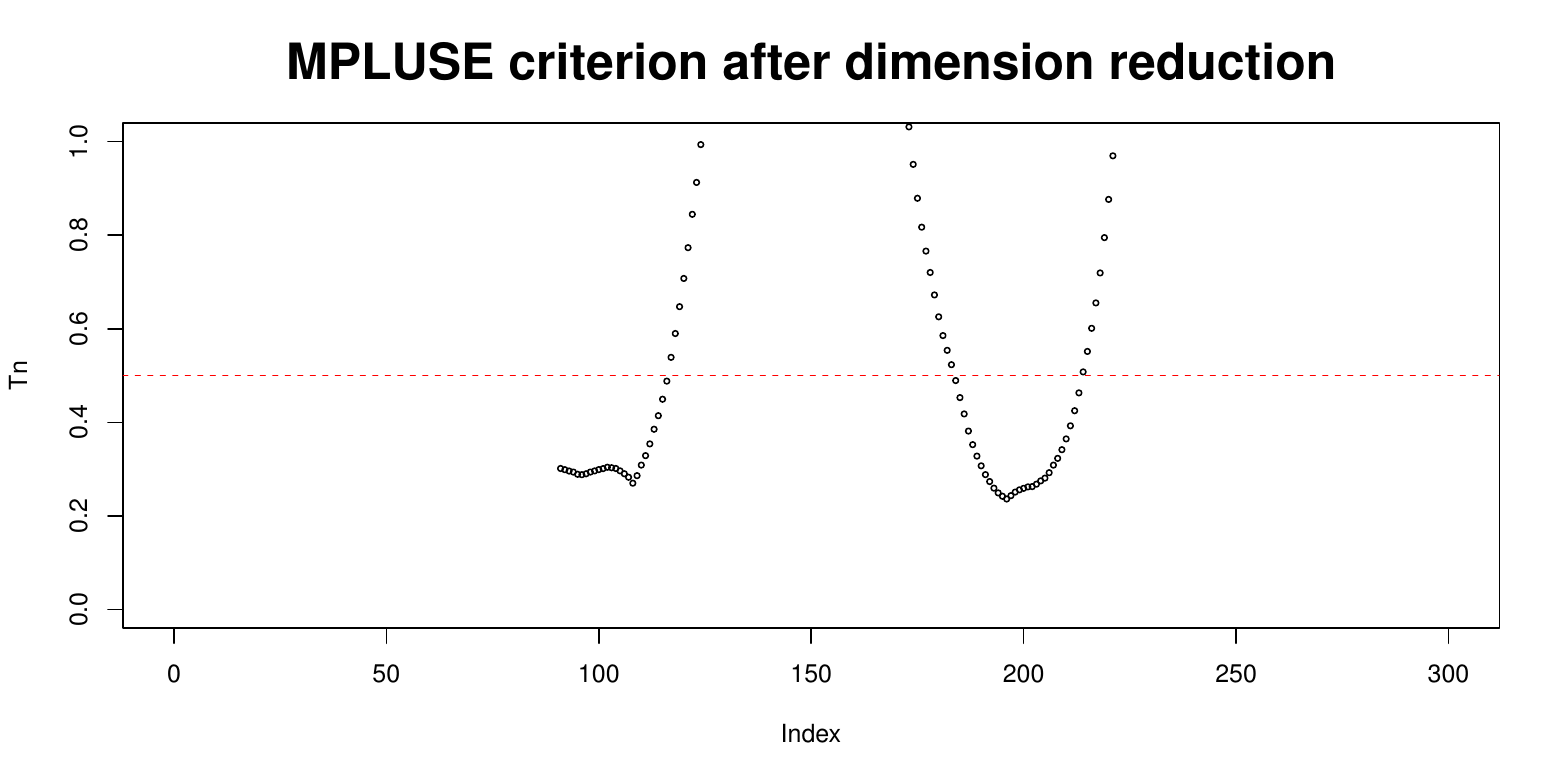}
	\caption{Plots before and after dimension reduction, and $S_{n}$ of MPULSE-ADS with $K=2$, $D_c=20$, and $u=0.1$.}
	\label{real_data1}
\end{figure}

To further investigate the mechanisms underlying the above phenomena, Figure~\ref{real_data1} presents four coordinated visualizations, including (1) the original functional data $\{Y_i(t)\}_{i=1}^n$, (2) the first dimension of the dimension-reduced data based on ADS, (3) the first dimension of the dimension-reduced data based on FPCA, and (4) the statistic $S_{n}$ of MPULSE-ADS with $K=2$, $D_c=20$, and $u=0.1$. The visualization demonstrates that our dimension reduction method preserves the essential change point structure while enabling  MPULSE-ADS to successfully detect the underlying structural break. Additionally, the jumps at change points are more pronounced in the ADS-based dimension-reduced data than in the FPCA-based results.

\section{Applications}

In this section, we implement both the ADS-based test and the adaptive functional change-point estimation algorithm to analyze financial and temperature datasets. The tuning parameters are configured identically to those employed in our simulation studies.

\subsection{CSI 300 Data}

Economic events trigger stock market fluctuations by altering market expectations and capital flows. For example, shifts in monetary policy or industrial policy frequently result in abrupt structural changes within the market. Identifying these change points allows for the precise mapping of risk transmission pathways, providing critical insights for investment strategies and policy assessments. 
The China Securities Index (CSI) 300, a pivotal stock index in China, serves as a core indicator reflecting Chinese economic trends due to its dynamic component adjustment mechanism and sectoral distribution. This study investigates change points in high-frequency CSI 300 data. For applications of high-frequency financial data in FDA, refer to \cite{Muller2011}.

\begin{figure}[htb!]
	\centering	\includegraphics[width=10cm,height=5cm]{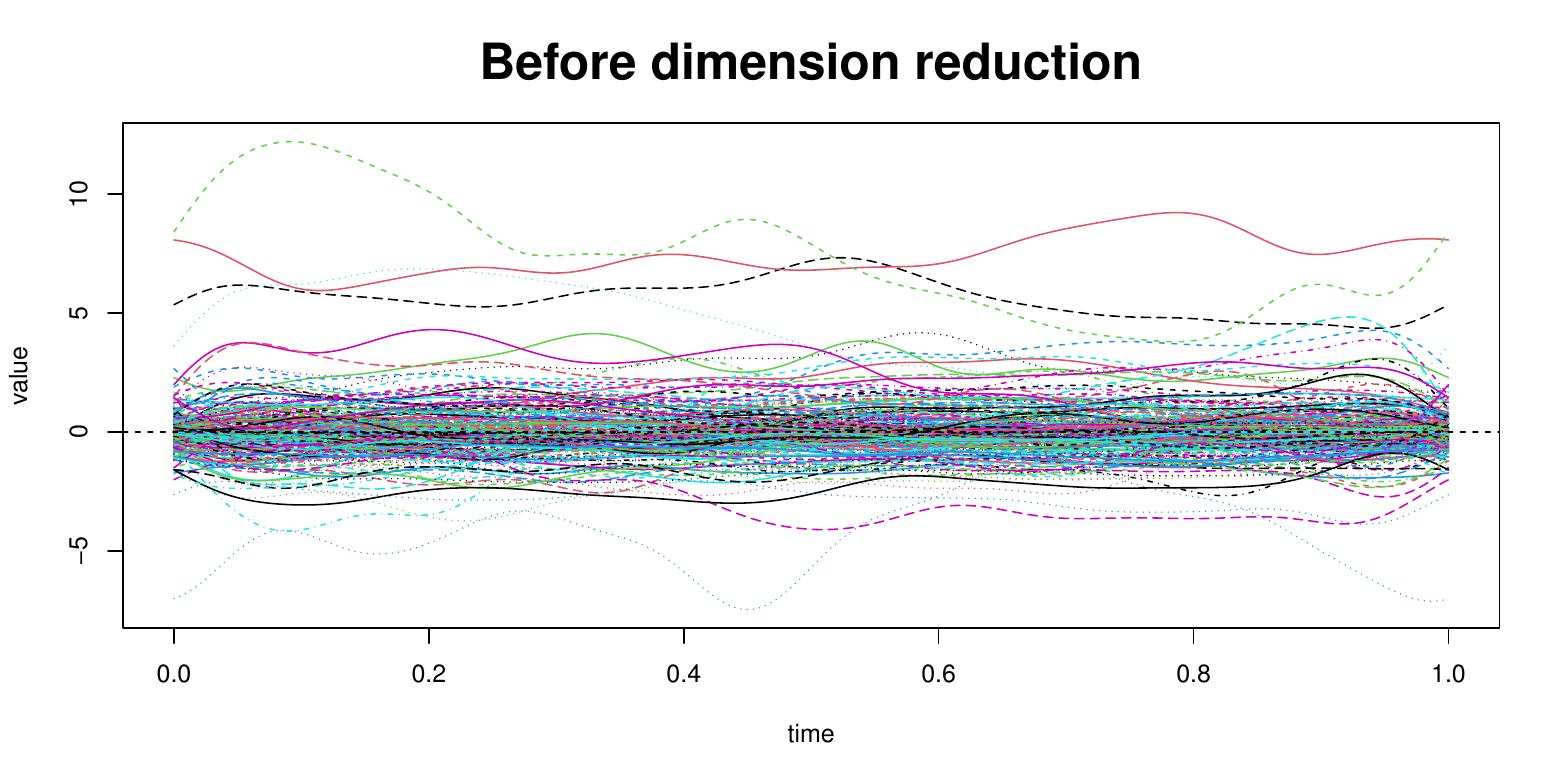}	\includegraphics[width=10cm,height=5cm]{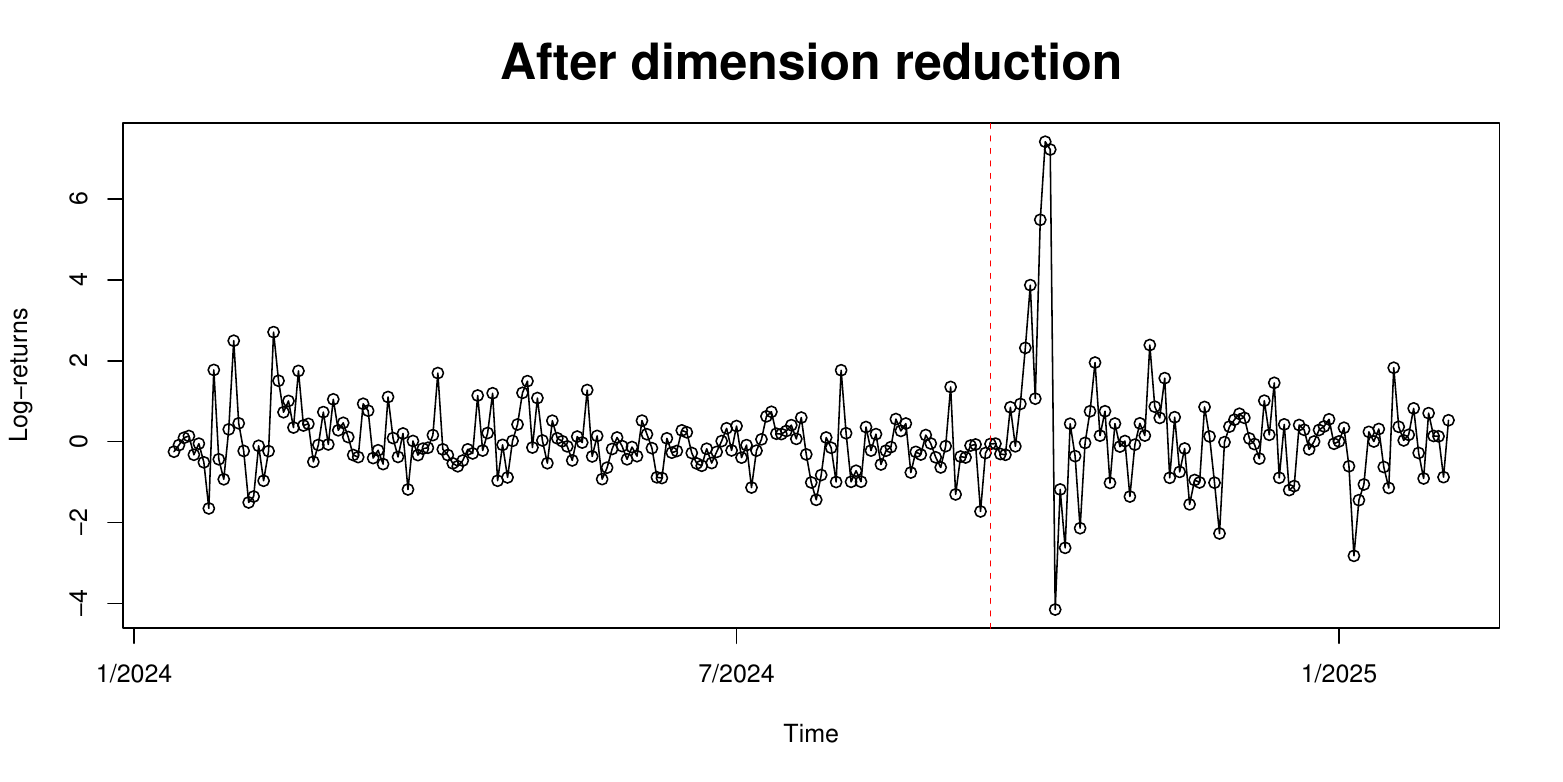}
	\caption{Figure before and after dimension reduction}
	\label{real_data4}
\end{figure}

The data in this study are sourced from the Wind Database (\url{https://www.wind.com.cn/portal/en/EDB/index.html}), covering the period from January 8, 2024, to February 6, 2025. For each trading day, closing prices are recorded in 1-minute intervals during two sessions: 9:30-11:30 AM and 1:00-3:00 PM, corresponding to a sampling frequency of 240 per day. We analyze the log-returns of these high-frequency observations. 
Daily log-returns are treated as functional objects, resulting in $n=257$ functional data points. We first apply our ADS-based test. The value of the test statistic $T_{2n}$ is   $11.3014$ with a $p$-value of approximately $0$, providing sufficient evidence to reject the null hypothesis at the  significance level $\alpha=0.05$.  
Applying our MPULSE-ADS criterion,  the estimated change point is located at $n=164$ corresponding to September 10, 2024. In September 2024, China’s stock market experienced a surge driven by aggressive policy easing, including RRR/interest rate cuts, equity-focused relending programs, and relaxed property regulations, while simultaneous Federal Reserve rate cuts amplified foreign capital inflows. Figure \ref{real_data4} presents the original functional data and the corresponding change point estimation after dimension reduction, effectively demonstrating the performance of our proposed method.

\subsection{Temperature Data}

\begin{figure}[htb!]
	\centering
	\includegraphics[width=10cm,height=5cm]{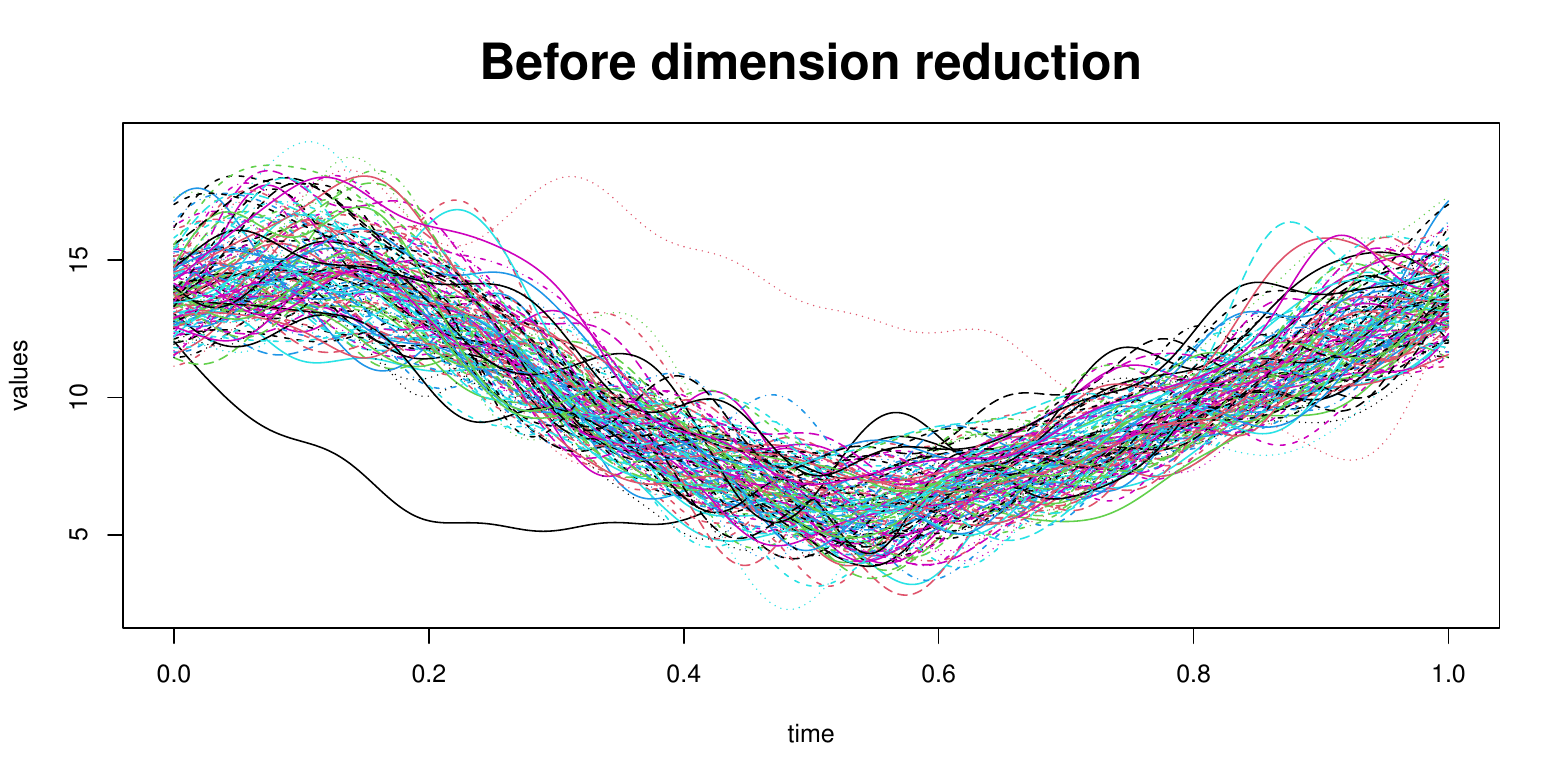}
	\includegraphics[width=10cm,height=5cm]{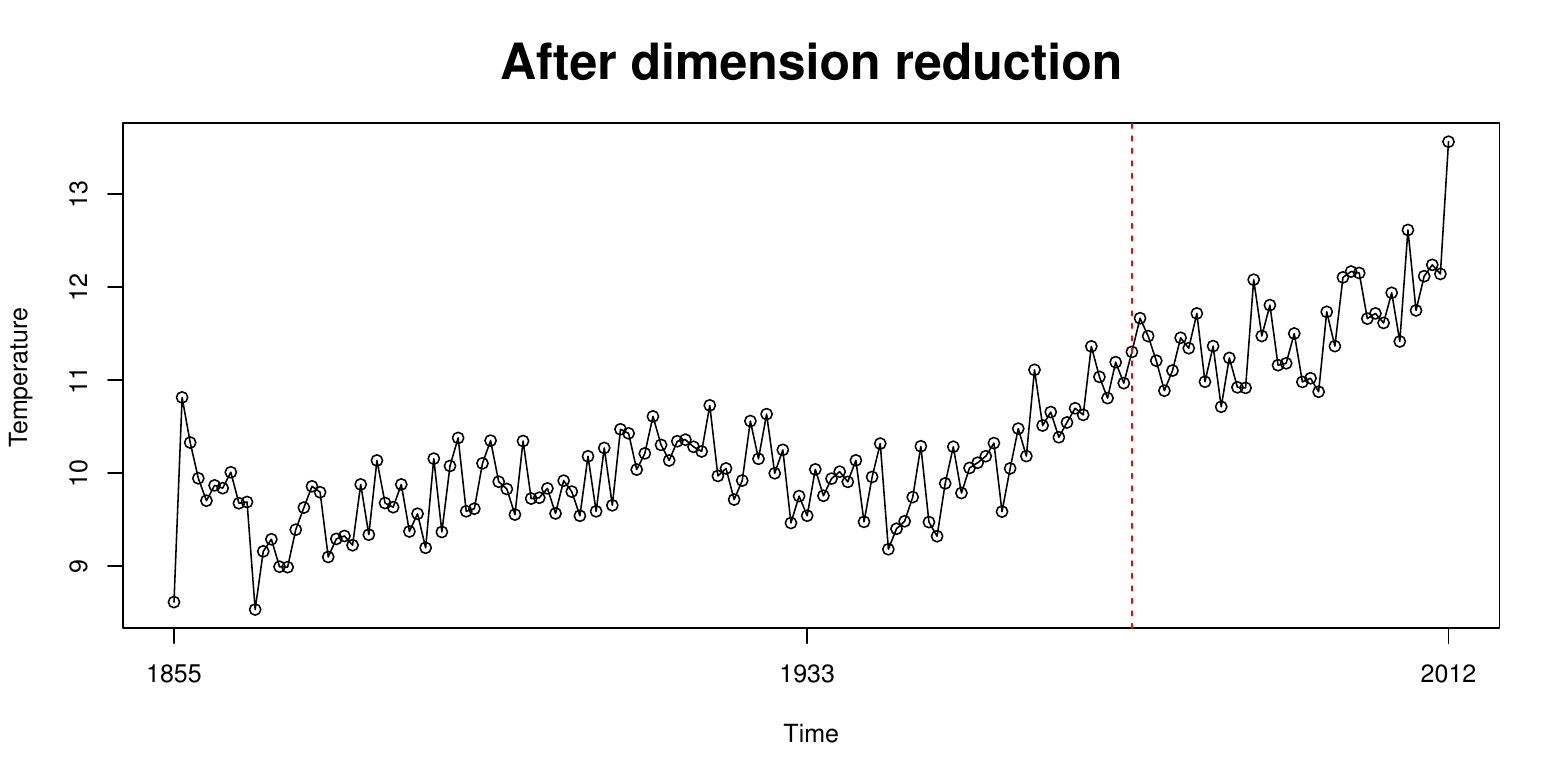}
	\caption{Figure before and after dimension reduction}
	\label{real_data3}
\end{figure}

In the context of global warming, change point detection in temperature data provides a key tool to identify distinct transitional phases of climate evolution.
This study investigates temperature variations in Melbourne, Australia, based on daily meteorological records spanning the period from 1855 to 2012, accessible via the Australian Bureau of Meteorology at
\url{www.bom.gov.au}. 
We model annual temperature profiles as functional objects, where each year is characterized by 365 daily measurements (or 366 for leap years). By applying a smoothing procedure with 21 Fourier basis functions, we convert the daily observations into functional objects, producing $n=158$ functional data points for further analysis.
Using our ADS-based test, the value of the  test statistic $T_{2n} $ is $35.5238$ with a $p$-value of approximately $0$. Thus, we have sufficient evidence to reject the null hypothesis at the significance level $\alpha=0.05$.  
Figure \ref{real_data3} illustrates the functional data and dimension-reduced data, highlighting significant structural shifts in Melbourne's temperature patterns.
Our proposed MPULSE-ADS criterion identifies a prominent change point at position 118, corresponding to the year 1973.  This detection aligns with historical climate milestones: starting in the early 1970s, the scientific community systematically confirmed the acceleration of global warming through meteorological evidence and raised formal alerts at pivotal international forums, including the 1972 Stockholm Conference on the Human Environment. This alignment underscores the methodological validity of our approach.

\section{Conclusion}

In this study, we introduce a novel concept called ADS for functional change point analysis. The proposed ADS framework effectively reduces infinite-dimensional functional data to finite-dimensional vector- or scalar-valued representations while preserving critical change-point information.
To identify this subspace, we develop an efficient estimator with solid theoretical guarantees for change-point information preservation. Leveraging the theoretical properties of the ADS target operator, we further construct an ADS-based test for change point testing. This test incorporates data splitting to optimize the selection of the projection direction.
Furthermore, we propose the MPULSE criterion to guarantee consistent change point estimation in the reduced-dimensional space. Our theoretical analysis rigorously establishes the asymptotic properties of all proposed methods, while numerical studies demonstrate their superior performance compared to FPCA-based approaches.

Despite these advancements, several challenges and opportunities for future research remain. First, while our framework primarily  focuses on changes in the mean structure of functional data, extending it to detect shifts in higher-order moments (e.g., covariance operators) or regression parameters in functional regression models could broaden its applicability. Second, as noted in \citep{zhao2020detecting}, identifying change points with small spacings remains a challenge. Third, the current framework is tailored for offline data analysis; adapting it to online settings with sequential data streams would enhance its utility in real-time monitoring scenarios. Addressing these challenges could pave the way for new research directions and further methodological innovations.

\newpage

\bibliographystyle{chicago}
\bibliography{arxiv}

\end{document}